\documentclass[epj,referee]{svjour}
\usepackage{amssymb}
\usepackage{cite}
\usepackage{exscale,axodraw,rotate}
\usepackage{epsfig}
\usepackage{color}
\begin{document}
%----------------------------------definitions
\def\bqa{\begin{eqnarray}}
\def\eqa{\end{eqnarray}}
\def\bq{\begin{equation}}
\def\eq{\end{equation}}
\newcommand{\nll}{\nonumber\\}
\newcommand{\sss}[1]{\scriptscriptstyle{#1}}
\newcommand{\ds }{\displaystyle}
\def\GF {G_{\sss F}}
\def\ml {m_{\ell}}
\def\mw {M_{\sss W}}
\def\gw {\Gamma_{\sss W}}
\def\mz {M_{\sss Z}}
\def\gz {\Gamma_{\sss Z}}
\def\mh {M_{\sss H}}
\def\stw{s_{\sss W}}
\def\ctw{c_{\sss W}}
\def\MSbar{\overline{MS}}
\newcommand{\GeV}{\unskip\,\mathrm{GeV}}
\newcommand{\MeV}{\unskip\,\mathrm{MeV}}
\def\hs{\hat s}
\def\hspr{\hat{s}'}
\def\qmo{Q_\ell}
\def\mmo{m_\ell}
\def\qup{Q_{u}}
\def\qqu{Q_{q}}
\def\mup{m_{u}}
\def\mtp{m_{t}}
\def\mbt{m_{b}}
\def\cmi{c_-}
\def\cpl{c_+}
\def\mqu{m_{q}}
\def\lnl{\ln_1}
\def\chic{\chi^*}
\def\order#1{{\mathcal O}\left(#1\right)}
\def\msbar{\overline{\tiny \mathrm{MS}}}
%----------------------------------------------
\def\koef{k_0}
\def\kk{k_1}
\def\kkk{k_2}
\def\stb{s_{tb}}
\def\smb{s_{b}}
\def\smt{s_{t}}
\def\mtb{m_{t+b}}
\def\thle{\theta_l}
%----------------------------------------------
\newcommand{\Litwo}{\mbox{${\rm{Li}}_{2}$}}
\newcommand{\als}{\alpha_{_S}}
\newcommand{\alsS}{\alpha^2_{_S}}
\def\wtp{\Gamma_t}
\def\mg{m_g}
\def\mwt {\widetilde{M}^2_{\sss{W}}}
\def\thle{\vartheta_{l}}
\def\cA{{\cal A}}
\def\SANC{{\tt SANC}~}
%----------------------------------------------
\def\Bd{B_{d}}
\def\Sd{S_{d}}
\def\ydl{y_{d1}}
\def\ydll{y_{d2}}
\def\Bl{B_{l}}
\def\Sl{S_{l}}
\def\yll{y_{d1}}
\def\ylll{y_{d2}}
\def\Qs{Q^2}
\def\mtpt{\tilde{m}_t}
\def\ieps{i\varepsilon}
%----------------------------------------------
\renewcommand{\Re}{\mbox{Re}\,}

\title{Electroweak Radiative Corrections to Single-top Production.}
\author{D. Bardin\inst{1} \email{bardin@nusun.jinr.ru}
%\thanks{This work is partly supported by Russian Foundation for Basic Research
%grant $N^{o}$ 10-02-01030.}
\and S. Bondarenko \inst{2}
\and L. Kalinovskaya \inst{1}
\and V. Kolesnikov \inst{1}
%\and G. Nanava\inst{3}\thanks{on leave from IHEP, TSU, Tbilisi, Georgia}
\and W. von Schlippe\inst{3}
}

\institute{Dzhelepov Laboratory for Nuclear Problems, JINR,    \\
           JINR,\ Dubna, \ 141980 \ \  Russia
\and
Bogoliubov Laboratory of  Theoretical Physics, JINR, \\
JINR,\ Dubna, \ 141980 \ \  Russia
%\and
%IFJ, \ PAN, \ 31-342  Krak\'ow \ \ Poland
%$^{3}$ Institute of Nuclear Physics, PAN,
%31-342  Krak\'ow, ul. Radzikowskiego 152, Poland,   \\
\and
Petersburg Nuclear Physics Institute, Gatchina, \ 188300 \ \  Russia
}

\date{Received: date / Revised version: date}

\abstract{
Radiative corrections to the single top $s$  and $t$ channel
production processes are revisited.
Complete one-loop electroweak corrections are calculated within
the \SANC system. 
New is a study of the regularisation of the top-legs associated infrared
divergences with aid of the complex mass of the top quark. A comparison
of these electroweak corrections with those computed by the conventional
method is presented both for top production and decays.
Standard FORM and FORTRAN  \SANC modules are created.
These modules are compiled into a package {\tt sanc\_cc\_v1.40}, which
may be downloaded from \SANC project homepages.
Numerous numerical results are presented at the partonic level with the aim
to demonstrate the correct working of modules. 
These modules are intended to be used in Monte Carlo generators for
single top production at the LHC.
Where possible, we compare our results with those existing in the literature;
in particular, a comprehensive comparison with results of the {\tt CompHEP} system 
is given.
\vspace*{-5mm}
\vspace*{-5mm}
}

\authorrunning{D. Bardin, et al.}
\titlerunning{Electroweak Radiative Corrections to Single-top...}

\maketitle

\clearpage

\section{Introduction}
%---------------------
In continuation of work on electroweak one-loop corrections to processes
involving the top quark~\cite{Arbuzov:2007ke,Bardin:2009wv}, we present here a
calculation of EW one-loop corrections to single top quark production
and top quark decay. To incorporate these processes in the \SANC
framework~\cite{Andonov:2004hi} is a natural extension of the previous work.

The interest in the study of single top production has been stated many times
(see {\it e.g.}~\cite{Schwienhorst:2008bs} and references therein), suffice it to say
that this is the only way of measuring the CKM matrix element $|V_{tb}|$
directly and thereby providing a sensitive test of the 3-generation scheme of the
Standard Model. Indeed, a significant deviation of $|V_{tb}|$ from the value
demanded by unitarity of the CKM matrix would be an indication of the presence
of a fourth generation of fundamental fermions.

Precision calculations of single top quark production have been done for
a number of years (see e.g.~\cite{Bernreuther:2008ju} and references therein)
but remain of continued interest. This is motivated on the one hand
by the observation of such events at the Tevatron D0~\cite{Abazov:2006gd,Abazov:2009ii}
and CDF~\cite{Aaltonen:2008sy,Aaltonen:2009jj}
and, on the other hand, by the need to prepare software to analyse single top quark events
at the LHC, running at 7 TeV.

Most of the theoretical work in single top production has been concerned with
higher order QCD corrections, leading to the development of Monte Carlo
generators, such as MC@NLO~\cite{Frixione:2005vw} or SingleTop~\cite{Boos:2006af},
incorporated in the standard LHC tools.

There are three channels of single top quark production: $t$ channel, $s$
channel and associated production. The $t$ channel process, $bq\to tq'$, where
$q$ and $q'$ are light quarks, is the channel known to have the largest cross
section~\cite{Beccaria:2006ir,Beccaria:2007tc,Beccaria:2008av,Mirabella:2008gj}.
 Calculations for the LHC at 14 TeV have led to the prediction of  this cross section being
more than an order of magnitude greater than that of the $s$ channel process
$qq'\to tb$ and about four times greater than the cross section of
associated production $qg\to tW$  (see {\it e.g.}~\cite{Kidonakis:2007ej}), and it
has been stated that the electroweak corrections are of comparable magnitude to
the QCD corrections \cite{Beccaria:2006ir}.

In this paper we calculate the electroweak corrections to the $s$ and $t$
channels of single top quark production and revisit the top quark
decay~\cite{Bardin:2009wv}.
The important new feature of the present work is the inclusion in the
calculations of the width $\Gamma_t$ of the top quark. This is a nontrivial step,
and we will show that the results are sensitive to $\Gamma_t$.

The paper is organized as follows.\\
In section 2 we present the covariant amplitudes for the processes under
consideration, working within the framework of the 
{\it multichannel approach} (see Ref~\cite{Bardin:2007wb}).
In section 3 we introduce the infrared regularisation by the complex mass of
the top quark, at first for the virtual QED corrections, and then for the hard
photon radiation. Numerical results are presented in section 4 and the
Conclusions and Outlook in section 5.

\section{Covariant Amplitude \label{Amplitudes}}
%-----------------------------------------------
In this section we proceed in the spirit of ``multi-channel approach'',
see Ref.~\cite{Bardin:2007wb}.
First, we consider annihilation into vacuum with all particles incoming.

\subsection{All particles incoming}
%----------------------------------
It is convenient to consider two vacuum diagrams:
\begin{figure}[!ht]
\begin{center}
   \begin{picture}(125,80)(210,10)
     \GOval(270,40)(34,5)(90){0.02}
     \ArrowLine(240,40)(200,70)
     \ArrowLine(200,10)(240,40)
     \ArrowLine(340,70)(300,40)
     \ArrowLine(300,40)(340,10)

     \ArrowLine(200,16)(220,32)
     \ArrowLine(200,64)(220,48)
     \ArrowLine(340,16)(320,32)
     \ArrowLine(340,64)(320,48)

     \Text(200,54)[]{\large $p_1$}
     \Text(200,28)[]{\large $p_2$}
     \Text(350,28)[]{\large $p_3$}
     \Text(350,54)[]{\large $p_4$}

     \Text(195,80)[]{\large$\bar{b}$}
     \Text(195,3)[]{\large $t$}
     \Text(350,5)[]{\large $\bar u$}
     \Text(350,80)[]{\large$d$}
   \end{picture}
\end{center}
\caption[The $t \bar{b} \bar{u} d\to 0$ processes]
        {The $t \bar{b} \bar{u} d\to 0$ processes}
\label{Diagram_tdec_vac}
\end{figure}
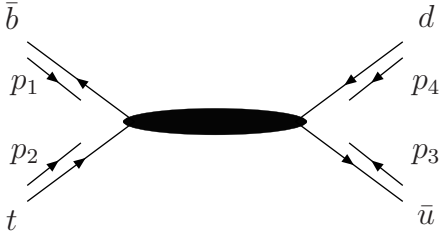

\noindent
and
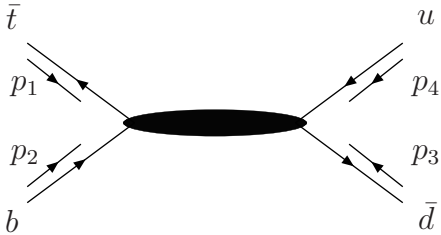
\begin{figure}[!ht]
\begin{center}
   \begin{picture}(125,80)(210,0)
     \GOval(270,40)(34,5)(90){0.02}
     \ArrowLine(240,40)(200,70)
     \ArrowLine(200,10)(240,40)
     \ArrowLine(340,70)(300,40)
     \ArrowLine(300,40)(340,10)

     \ArrowLine(200,16)(220,32)
     \ArrowLine(200,64)(220,48)
     \ArrowLine(340,16)(320,32)
     \ArrowLine(340,64)(320,48)

     \Text(200,54)[]{\large $p_1$}
     \Text(200,28)[]{\large $p_2$}
     \Text(350,28)[]{\large $p_3$}
     \Text(350,54)[]{\large $p_4$}

     \Text(195,80)[]{\large$\bar{t}$}
     \Text(195,3)[]{\large $b$}
     \Text(350,5)[]{\large $\bar d$}
     \Text(350,80)[]{\large$u$}
   \end{picture}
\end{center}
\caption[The $\bar{t} b  u \bar{d} \to 0$ processes]
        {The $\bar{t} b  u \bar{d} \to 0$ processes}
\label{Diagram_atdec_vac}
\end{figure}

Here black ovals symbolically denote all one-loop insertions to the
corresponding tree diagrams.

Covariant Amplitudes (CA) of Figs.~\ref{Diagram_tdec_vac} and~\ref{Diagram_atdec_vac} are
characterized by four different structures and scalar Form Factors (FF)
if the mass of the light quarks is neglected the and $b$ quark mass
is not neglected.
Omitting Dirac spinors, one may write a common expression for this CA
of the process in terms of scalar form factors, ${\cal F}_{\sss{LL}}(s,t)$,
see~\cite{Andonov:2004hi}:
\bqa
{\cal A} &=&i\,e^2\,\frac{d_{\sss W}(s)}{4}\Big[
    \gamma_\mu \left( 1+\gamma_5 \right) \otimes \gamma_\mu \left( 1+\gamma_5 \right)
                       {\cal F}_{\sss{LL}}(s,t)\,
\nll
&&+ \gamma_\mu \left( 1-\gamma_5 \right) \otimes \gamma_\mu \left( 1+\gamma_5 \right)
                       {\cal F}_{\sss{RL}}(s,t)\,
\nll
&&+ \left( 1+\gamma_5 \right) \otimes \gamma_\mu \left( 1+\gamma_5 \right)(-i D_\mu)
                       {\cal F}_{\sss{LD}}(s,t)\,
\nll
&&+ \left( 1-\gamma_5 \right) \otimes \gamma_\mu \left( 1+\gamma_5 \right)(-i D_\mu)
                       {\cal F}_{\sss{RD}}(s,t)\Big],
\eqa
%---
with
\bq
D_\mu=(p_1-p_2)_\mu\,,
\eq
%--
and 4-momentum conservation reads
\bq
p_1+p_2+p_3+p_4=0\,,
\eq
%--
the invariants are defined by
\bq
s=-(p_1+p_2)^2,\quad t=-(p_2+p_3)^2,
\eq
%--
and
\bq
d_{\sss W}(s)=\frac{V}{2\stw^2}\,\frac{1}{s-\mw^2+i\mw\Gamma_{\sss W}}\,.
\eq
%--
Here $V=V_{tb}V_{ud}$ is the relevant product of CKM matrix elements
and the symbol $\otimes$ means, considering the factor of ${\cal{F}}_{LL}$ in
diagram Fig.~\ref{Diagram_tdec_vac},
\begin{eqnarray*}
&&\gamma_\mu\gamma_+\otimes\gamma_\mu\gamma_+
=\bar{u}(p_2) \gamma_\mu\gamma_+v(p_1)\bar{v}(p_3) \gamma_\mu\gamma_+u(p_4)\\
&& \mbox{with}\qquad \gamma_+ = 1+\gamma_5
\end{eqnarray*}
and appropriate changes in the other cases.
The scalar form factors ${\cal{F}}$ are labeled according to
their structures, see \cite{Andonov:2004hi}.

\subsection{Conversion to top decay}
%-----------------------------------
The CA for the decay $t(p_2) \to b(p_1) + u(p_3) + \bar{d}(p_4)$
is derived from Fig.~\ref{Diagram_tdec_vac} by
the following 4-momentum replacement:
\[
  \begin{array}{llllllll}
    & p_1 & \to & -p_1,\quad&\quad  p_3 & \to & -p_3, \\[-1mm]
    & p_2 & \to &~~p_2,\quad&\quad  p_4 & \to & -p_4. \\
  \end{array}
\]

For the decay $\bar{t}(p_1) \to \bar{b}(p_2) + d(p_3) + \bar{u}(p_4)$ it
is more transparent to make the replacement from Fig.~\ref{Diagram_atdec_vac}:
\[
  \begin{array}{llllllll}
    & p_1 & \to &~~p_1,\quad&\quad  p_3 & \to & -p_3, \\[-1mm]
    & p_2 & \to & -p_2,\quad&\quad  p_4 & \to & -p_4. \\
  \end{array}
\]

As a result one has two decay diagrams, schematically representing one-loop EW corrections:

\begin{figure}[!ht]
\begin{center}
   \begin{picture}(125,80)(210,10)
     \GOval(270,40)(34,5)(90){0.02}
     \ArrowLine(240,40)(300,58)
     \ArrowLine(200,10)(240,40)
     \ArrowLine(340,70)(300,40)
     \ArrowLine(300,40)(340,10)

     \ArrowLine(200,16)(220,32)
     \ArrowLine(250,48)(275,57)
     \ArrowLine(320,32)(340,16)
     \ArrowLine(320,48)(340,64)

     \Text(260,65)[]{\large $p_1$}
     \Text(200,28)[]{\large $p_2$}
     \Text(350,28)[]{\large $p_3$}
     \Text(350,54)[]{\large $p_4$}

     \Text(307,60)[]{\large$b$}
     \Text(195,3)[]{\large $t$}
     \Text(350,5)[]{\large $u$}
     \Text(350,80)[]{\large$\bar{d}$}
   \end{picture}
\end{center}
\caption[The $t\to b u \bar{d}$ processes]
        {The $t\to b u \bar{d}$ processes}
\label{Diagram_tdec}
\end{figure}

\begin{figure}[!ht]
\begin{center}
   \begin{picture}(125,80)(210,10)
     \GOval(270,40)(34,5)(90){0.02}

     \ArrowLine(240,40)(200,70)
     \ArrowLine(340,70)(300,40)
     \ArrowLine(300,40)(340,10)
     \ArrowLine(240,40)(300,20)
%p_2
     \ArrowLine(258,28)(286,19)

     \ArrowLine(200,64)(220,48)
     \ArrowLine(320,32)(340,16)
     \ArrowLine(320,48)(340,64)

     \Text(200,54)[]{\large $p_1$}
     \Text(350,28)[]{\large $p_3$}
     \Text(350,54)[]{\large $p_4$}
     \Text(260,16)[]{\large $p_2$}

     \Text(195,80)[]{\large$\bar{t}$}
     \Text(350,5)[]{\large $ d$}
     \Text(350,80)[]{\large$\bar u$}
     \Text(307,10)[]{\large$\bar{b}$}

   \end{picture}
\end{center}
\caption[The $\bar{t}\to \bar{b} \bar{u} d $ processes]
        {The $\bar{t}\to \bar{b} \bar{u} d $ processes}
\label{Diagram_atdec}
\end{figure}

From the diagram of Fig.~\ref{Diagram_tdec} the four Helicity Amplitudes
(HA) can be derived by the standard techniques used in {\tt SANC},
see Eqs.~(47-50) of Ref.~\cite{Andonov:2004hi}.

For the case of the process of diagram of Fig.~\ref{Diagram_atdec} the
HA are similar, although not identical.
For their exact expressions see the relevant module in the {\tt SANC} tree.
%{\bf EW $\to$ Processes $\to$ 4 legs $\to$ Charged Current $\to$ t$\to$ b~u~d (HA)}.

\subsection{Conversion to $s$ channel}
%-----------------------------------
The CA for the $s$ channel single-top production processes
$\bar{u}(p_1)+d(p_2)\to b(p_3)+\bar{t}(p_4)$ is obtained from Fig.~\ref{Diagram_tdec_vac}
by the following 4-momentum replacement:
\[
  \begin{array}{llllllll}
    & p_1 & \to & -p_3,\quad&\quad  p_3 & \to &~~p_1, \\[-1mm]
    & p_2 & \to & -p_4,\quad&\quad  p_4 & \to &~~p_2, \\
  \end{array}
\]

\noindent
leading to the diagram

\begin{figure}[!ht]
\begin{center}
   \begin{picture}(125,80)(210,0)
     \GOval(270,40)(34,5)(90){0.02}
     \ArrowLine(240,40)(200,70)
     \ArrowLine(200,10)(240,40)
     \ArrowLine(340,70)(300,40)
     \ArrowLine(300,40)(340,10)

     \ArrowLine(200,16)(220,32)
     \ArrowLine(200,64)(220,48)
     \ArrowLine(320,32)(340,16)
     \ArrowLine(320,48)(340,64)

     \Text(200,54)[]{\large $p_1$}
     \Text(200,28)[]{\large $p_2$}
     \Text(350,28)[]{\large $p_3$}
     \Text(350,54)[]{\large $p_4$}

     \Text(195,80)[]{\large$\bar{u}$}
     \Text(195,3)[]{\large $d$}
     \Text(350,5)[]{\large $b$}
     \Text(350,80)[]{\large$\bar t$}
   \end{picture}
\end{center}
\caption[The $\bar{u} d \to \bar{t} b$ process]
        {The $\bar{u} d \to \bar{t} b$ process}
\label{Diagram_schannel_mi}
\end{figure}

\noindent
For the processes $\bar{d}(p_1)+u(p_2)\to t(p_3)+\bar{b}(p_4)$ from
Fig.~\ref{Diagram_atdec_vac} the conversion
\[
  \begin{array}{llllllll}
    & p_1 & \to & -p_3,\quad&\quad  p_3 & \to &~~p_1, \\[-1mm]
    & p_2 & \to & -p_4,\quad&\quad  p_4 & \to &~~p_2, \\
  \end{array}
\]

\noindent
leads to the diagram of Fig.~\ref{Diagram_schannel_pl}.

\begin{figure}[!ht]
\begin{center}
   \begin{picture}(125,80)(210,0)
     \GOval(270,40)(34,5)(90){0.02}
     \ArrowLine(240,40)(200,70)
     \ArrowLine(200,10)(240,40)
     \ArrowLine(340,70)(300,40)
     \ArrowLine(300,40)(340,10)

     \ArrowLine(200,16)(220,32)
     \ArrowLine(200,64)(220,48)
     \ArrowLine(320,32)(340,16)
     \ArrowLine(320,48)(340,64)

     \Text(200,54)[]{\large $p_1$}
     \Text(200,28)[]{\large $p_2$}
     \Text(350,28)[]{\large $p_3$}
     \Text(350,54)[]{\large $p_4$}

     \Text(195,80)[]{\large$\bar{d}$}
     \Text(195,3)[]{\large $u$}
     \Text(350,5)[]{\large $t$}
     \Text(350,80)[]{\large$\bar b$}
   \end{picture}
\end{center}
\caption[The $\bar{d} u\to t\bar{b}$ processes]
        {The $\bar{d} u\to t\bar{b}$ processes}
\label{Diagram_schannel_pl}
\end{figure}

For the corresponding HA one can get rather compact expressions.

$\bullet$ Helicity amplitude for $\bar{u}d\to b\bar{t}$
%\_W^-
\bqa
\cal{H}_{+---}&=&  k_0 \sin\vartheta\bigl\{(\mtp k_1+\mbt k_2) {\cal F}_{LL}
\nll &&
+(\mtp k_2+\mbt k_1) {\cal F}_{LR}
\nll &&
                 -P^{+}P^{-}(k_2 {\cal F}_{LD}-k_1 {\cal F}_{RD}) \bigr\},
\nll
\cal{H}_{+-++}&=& k_0\sin\vartheta\bigl\{(\mtp k_2+\mbt k_1) {\cal F}_{LL}
\nll &&
+(\mtp k_1+\mbt k_2) {\cal F}_{LR}
\nll &&
                 +P^{+}P^{-}(k_1 {\cal F}_{LD}-k_2{\cal F}_{RD}) \bigr\},
\nll
\cal{H}_{+--+} &=& k_0\sqrt{s} \cpl \bigl(k_1{\cal F}_{LL}+k_2{\cal F}_{LR}\bigr),
\nll
\cal{H}_{+-+-} &=& k_0\sqrt{s} \cmi \bigl(k_2{\cal F}_{LL}+k_1{\cal F}_{LR}\bigr).
\eqa

$\bullet$ Helicity amplitude for $\bar{d}u\to t\bar{b}$
%\_W^+
\bqa
\cal{H}_{+---}&=& k_0 \sin\vartheta\bigl\{(\mtp k_2+\mbt k_1) {\cal F}_{LL}
\nll &&
+(\mtp k_1+\mbt k_2) {\cal F}_{LR}
\nll &&
                 -P^{+}P^{-}(k_2 {\cal F}_{LD}-k_1 {\cal F}_{RD}) \bigr\},
\nll
\cal{H}_{+-++}&=& k_0 \sin\vartheta\bigl\{(\mtp k_1+\mbt k_2) {\cal F}_{LL}
\nll &&
+(\mtp k_2+\mbt k_1) {\cal F}_{LR}
\nll &&
                 +P^{+}P^{-}(k_1 {\cal F}_{LD}-k_2 {\cal F}_{RD}) \bigr\},
\nll
\cal{H}_{+-+-}&=& k_0\sqrt{s} \cmi \bigl(k_2{\cal F}_{LL}+k_1{\cal F}_{LR}\bigr),
\nll
\cal{H}_{+--+}&=& k_0\sqrt{s} \cpl \bigl(k_1{\cal F}_{LL}+k_2{\cal F}_{LR}\bigr).
\eqa
Here
\bqa
 k_0&=&-\frac{1}{2}V\chi(\mw^2,s),
\nll
\chi(\mw^2,s)&=&\frac{s}{2s_W^2}\frac{1}{s-M_W^2+iM_W\gamma_W}
\nll
 k_{1,2}&=& P^{-} \pm P^{+},
\nll
 P^{\pm}&\equiv&\sqrt{s-(\mtp\pm\mbt)^2}\,,
\nll
c_\pm&=&1\pm\cos\vartheta\,
\eqa
and $\vartheta$ is always the angle $\angle(\vec{p}_1,\vec{p}_3)$.

\subsection{Conversion to $t$ channel processes}
%-----------------------------------------------
In the CAs for the $t$ channel single-top production processes
$\bar{b}(p_1)+\bar{u}(p_2)\to\bar{d}(p_3)+\bar{t}(p_4)$ and
$\bar{b}(p_1)+d(p_2)\to u(p_3)+\bar{t}(p_4)$
it is convenient to make the replacement ``in pairs''.
From Fig.~\ref{Diagram_tdec_vac} one may perform two 4-momentum replacements:
\[
  \begin{array}{llllllll}
    &\hspace{7mm} p_1 & \to &~~p_1,\hspace{23mm} & p_1 & \to &~~p_1,\\[-1mm]
    &\hspace{7mm} p_2 & \to & -p_4,\hspace{23mm} & p_2 & \to & -p_4,\\[-1mm]
    &\hspace{7mm} p_3 & \to &~~p_2,\hspace{23mm} & p_3 & \to & -p_3,\\[-1mm]
    &\hspace{7mm} p_4 & \to & -p_3,\hspace{23mm} & p_4 & \to &~~p_2,\\
  \end{array}
\]

\noindent
which give rise to two different physical $t$ channel processes, described
by two symbolic diagrams:
\begin{figure}[!ht]
\begin{center}
\begin{tabular}{cc}
  \begin{picture}(125,80)(210,20)
    \GOval(270,40)(34,5)(0){0.02}
    \ArrowLine(270,67)(230,97)
    \ArrowLine(310,97)(270,67)
    \ArrowLine(230,-17)(270,13)
    \ArrowLine(270,13)(310,-17)
    \ArrowLine(240,-20)(260,-4)
    \ArrowLine(240,100)(260,84)
    \ArrowLine(280,84)(300,100)
    \ArrowLine(280,-4)(300,-20)
    \Text(230,80)[]{\large $\bar b$}
    \Text(230,5)[]{\large $\bar u$}
    \Text(310,5)[]{\large $\bar d$}
    \Text(310,80)[]{\large $\bar t$}
    \Text(260,100)[]{\large $p_1$}
    \Text(285,100)[]{\large $p_4$}
    \Text(285,-20)[]{\large $p_3$}
    \Text(260,-20)[]{\large $p_2$}
  \end{picture}
\hspace*{-5mm}
&
\hspace*{-5mm}
  \begin{picture}(125,80)(210,20)
    \GOval(270,40)(34,5)(0){0.02}
    \ArrowLine(270,67)(230,97)
    \ArrowLine(310,97)(270,67)
    \ArrowLine(230,-17)(270,13)
    \ArrowLine(270,13)(310,-17)
    \ArrowLine(240,-20)(260,-4)
    \ArrowLine(240,100)(260,84)
    \ArrowLine(280,84)(300,100)
    \ArrowLine(280,-4)(300,-20)
    \Text(230,80)[]{\large $\bar b$}
    \Text(230,5)[]{\large $d$}
    \Text(310,5)[]{\large $u$}
    \Text(310,80)[]{\large $\bar t$}
    \Text(260,100)[]{\large $p_1$}
    \Text(285,100)[]{\large $p_4$}
    \Text(285,-20)[]{\large $p_3$}
    \Text(260,-20)[]{\large $p_2$}
  \end{picture}
\end{tabular}
\end{center}
\vspace*{15mm}
\caption[The $\bar{b}\bar{u}\to\bar{t}\bar{d}$ and $\bar{b}d\to\bar{t}u$ processes]
        {The $\bar{b}\bar{u}\to\bar{t}\bar{d}$ and $\bar{b}d\to\bar{t}u$ processes}
\end{figure}
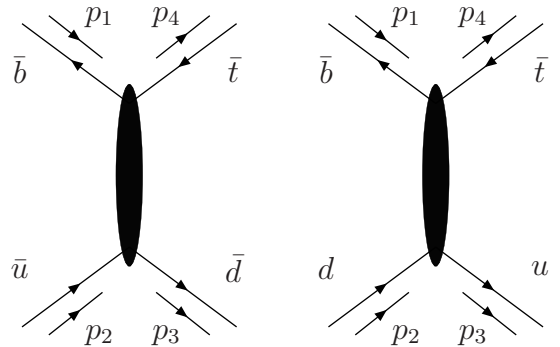

For the processes $b(p_1)+u(p_2)\to t(p_4)+d(p_3)$ and
 $b(p_1)+\bar{d}(p_2)\to t(p4)+\bar{u}(p_3)$,
the pair of replacements from Fig.~\ref{Diagram_atdec_vac},
\[
  \begin{array}{llllllll}
    &\hspace{7mm} p_1 & \to & -p_4,\hspace{23mm} & p_1 & \to & -p_4,\\[-1mm]
    &\hspace{7mm} p_2 & \to &~~p_1,\hspace{23mm} & p_2 & \to &~~p_1,\\[-1mm]
    &\hspace{7mm} p_3 & \to & -p_3,\hspace{23mm} & p_3 & \to &~~p_2,\\[-1mm]
    &\hspace{7mm} p_4 & \to &~~p_2,\hspace{23mm} & p_4 & \to & -p_3,\\
  \end{array}
\]

\noindent
gives the corresponding pair of symbolic diagrams for the two remaining
 $t$ channel processes:

\begin{figure}[!ht]
\begin{center}
\begin{tabular}{cc}
  \begin{picture}(125,80)(210,30)
    \GOval(270,40)(34,5)(0){0.02}
    \ArrowLine(230,97)(270,67)
    \ArrowLine(270,67)(310,97)
    \ArrowLine(230,-17)(270,13)
    \ArrowLine(270,13)(310,-17)
    \ArrowLine(240,-20)(260,-4)
    \ArrowLine(240,100)(260,84)
    \ArrowLine(280,84)(300,100)
    \ArrowLine(280,-4)(300,-20)
    \Text(230,80)[]{\large $b$}
    \Text(230,5)[]{\large $u$}
    \Text(310,5)[]{\large $d$}
    \Text(310,80)[]{\large $t$}
    \Text(260,100)[]{\large $p_1$}
    \Text(285,100)[]{\large $p_4$}
    \Text(285,-20)[]{\large $p_3$}
    \Text(260,-20)[]{\large $p_2$}
  \end{picture}
\hspace*{-5mm}
&
\hspace*{-5mm}
  \begin{picture}(125,80)(210,30)
    \GOval(270,40)(34,5)(0){0.02}
    \ArrowLine(230,97)(270,67)
    \ArrowLine(270,67)(310,97)
    \ArrowLine(230,-17)(270,13)
    \ArrowLine(270,13)(310,-17)
    \ArrowLine(240,-20)(260,-4)
    \ArrowLine(240,100)(260,84)
    \ArrowLine(280,84)(300,100)
    \ArrowLine(280,-4)(300,-20)
    \Text(230,80)[]{\large $b$}
    \Text(230,5)[]{\large $\bar d$}
    \Text(310,5)[]{\large $\bar u$}
    \Text(310,80)[]{\large $t$}
    \Text(260,100)[]{\large $p_1$}
    \Text(285,100)[]{\large $p_4$}
    \Text(285,-20)[]{\large $p_3$}
    \Text(260,-20)[]{\large $p_2$}
  \end{picture}
\end{tabular}
\end{center}
\vspace*{2cm}
\caption[The $bu\to td$ and $b\bar{d}\to t\bar{u}$ processes]
        {The $bu\to td$ and $b\bar{d}\to t\bar{u}$ processes}
\label{Diagrammsl}
\end{figure}
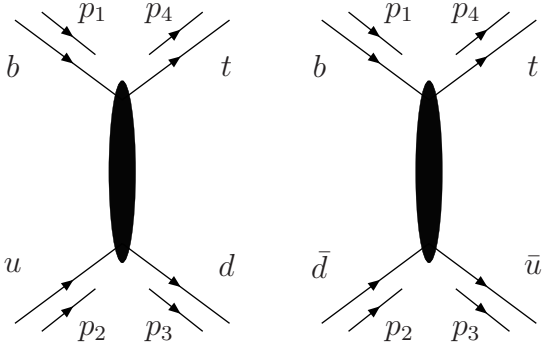

The HAs are all different for the four types of processes.
They are listed below in another sequence than they were presented
diagrammatically above. First, we give the
more complicated HAs for 4-particle and 4-antiparticle processes:

$\bullet$ {Helicity amplitudes for $ub\to td$
%\_1\_13
%-------
\bqa
\label{ubtd1_13}
\cal{H}_{----} &=& \koef \Bigl\{
       N^{+}\sqrt{s}\cpl\Bigl[ 2 \mtb {\cal F}_{LL}
\nll &&
+\frac{\mtp}{s}\bigl(\cmi\smb{\cal F}_{RL}-\kkk{\cal F}_{LD}\bigr)
-\kk {\cal F}_{RD}\Bigr]
\nll &&
      -N^{-}\cmi \Bigl[2\stb {\cal F}_{LL}
\nll &&
-\mtp\Bigl(\frac{\kkk}{s}{\cal F}_{RL}+\cpl\smb\left({\cal F}_{LD}-{\cal F}_{RD}\right)\Bigr)
\Bigr]\Bigr\},
\nll
\cal{H}_{---+} &=& \koef \sin\vartheta \Bigl\{
       N^{+}\cpl\smb\Bigl[ {\cal F}_{RL}
\nll &&
     +\mtp \left( {\cal F}_{LD}-{\cal F}_{RD}\right)\Bigr]
\nll &&
     +N^{-}\sqrt{s}\Bigl[\frac{\kkk}{s}{\cal F}_{RL}
\nll &&
      -\kk{\cal F}_{LD}-\frac{\mtp\kkk}{s}{\cal F}_{RD}\Bigr]\Bigr\},
\nll
\cal{H}_{+---} &=& \koef \sin\vartheta \Bigl\{
      -N^{+} \Bigl[2\stb {\cal F}_{LL}
\nll &&
        -\mtp \Bigl(\frac{\kkk}{s} {\cal F}_{RL}
        +\cpl\smb ({\cal F}_{LD}-{\cal F}_{RD})\Bigr) \Bigr]
\nll &&
      -N^{-} \sqrt{s} \Bigl[ 2\mtb {\cal F}_{LL}
\nll &&
       +\frac{\mtp}{s}\bigl(\cmi\smb{\cal F}_{RL}-\kkk {\cal F}_{LD}\bigr)
       -\kk {\cal F}_{RD}\Bigr]\Bigr\},
\nll
\cal{H}_{+--+} &=& \koef \cpl \Bigl\{
       N^{+} \sqrt{s} \left( \frac{\kkk}{s} {\cal F}_{RL}\right.
\nll &&
\left. -\kk {\cal F}_{LD}-\frac{\mtp\kkk}{s} {\cal F}_{RD} \right)
\nll &&
      -N^{-} \cmi\smb\bigl[ {\cal F}_{RL}+\mtp ({\cal F}_{LD}-{\cal F}_{RD})\bigr]\Bigr\}.\;
\eqa
Specific to Eq.~(\ref{ubtd1_13}) is the following notation:
\bqa
 \koef&=& V\chi(\mw^2,s)N^2 N^{-}_{(2,3)}, \nll
 \kk  &=& 2\stb-\cpl\smb,\nll
 \kkk &=& 2\stb\mbt+\cpl\mtp \smb.
\eqa
The other notation common to Eqs.~(\ref{ubtd1_13}) and (\ref{ubtd2_13}) is
\bqa
 \stb &=& s + \mtp\mbt,\qquad
 \mtb\;=\; \mtp+\mbt, \nll
 \smt &=& s - \mtp^2, \qquad\qquad\;\, \smb \;=\; s - \mbt^2, \nll
 c_\pm&=& 1\pm \cos\vartheta,
\eqa
 and
\bqa
N    &=&\sqrt{\frac{2s}{\stb^2\cmi + s\mtp^2\cpl}}\,,\nll
N^{-}_{(2,3)}&\equiv&N^{-}_{(3,2)}\;=\;\sqrt{\frac{\smt\smb}{2}},\nll
N^{+}&=&\frac{\mtb}{\sqrt{2}}\,,\quad
N^{-}\;=\;\frac{\stb}{\sqrt{2s}}\,.
\eqa

$\bullet$ Helicity amplitudes for $\bar{b}\bar{u}\to\bar{t}\bar{d}$
%\_2\_13
%-------
\bqa
\label{ubtd2_13}
\cal{H}_{----} &=& \koef \Bigl\{
       -N^{+}\sqrt{s}\cpl\Bigl[\frac{\kkk}{s} {\cal F}_{RL}
\nll &&
        -\bigl(2\mtb\mtp+\cmi\smt\bigr){\cal F}_{LD}-\frac{\mbt\kkk}{s}{\cal F}_{RD}\Bigr]
\nll &&
       -N^{-} \cmi \Bigl[2 \stb {\cal F}_{LL}
\nll &&
-\mbt\Bigl(\frac{\kkk}{s} {\cal F}_{RL}
+\cpl\smt ({\cal F}_{LD}-{\cal F}_{RD})\Bigl)\Bigr]\Bigr\},
\nll
\cal{H}_{---+} &=& \koef\sin\vartheta \Bigl\{
        N^{+}\Bigl[2 \stb {\cal F}_{LL}
\nll &&
- \mbt\Bigl(\frac{\kkk}{s} {\cal F}_{RL}
 +\cpl\smt({\cal F}_{LD}-{\cal F}_{RD})\Bigr)\Bigr]
\nll &&
       -N^{-} \sqrt{s}\Bigl[\frac{\kkk}{s} {\cal F}_{RL}
\nll &&
       -\kk{\cal F}_{LD}-\frac{\mbt\kkk}{s}{\cal F}_{RD}\Bigr]\Bigr\},
\nll
\cal{H}_{+---} &=& \koef\sin\vartheta \Bigl\{
       -N^{+} \cpl \smt \bigr[{\cal F}_{RL}
\nll &&
+\mbt \left( {\cal F}_{LD}-{\cal F}_{RD}\right)\bigr]
\nll &&
       +N^{-}\sqrt{s}\Bigr[2\mtb {\cal F}_{LL}
\nll &&
       +\frac{\mbt}{s}\bigl(\cmi\smt{\cal F}_{RL}-\kkk{\cal F}_{LD}\bigr)-\kk {\cal F}_{RD}
        \Bigr]\Bigr\},
\nll
\cal{H}_{+--+} &=& \koef \cpl \Bigl\{
       -N^{+} \sqrt{s} \Bigl[2\mtb {\cal F}_{LL}
\nll &&
+\frac{\mbt}{s}\bigl(\cmi\smt {\cal F}_{RL}+\kkk{\cal F}_{LD}\bigr)
     +\kk {\cal F}_{RD}\Bigr]
\nll &&
      -N^{-} \cmi \smt\bigl[{\cal F}_{RL}
+\mbt ({\cal F}_{LD}-{\cal F}_{RD})\bigr] \Bigr\}.\;
\eqa
Notation specific to Eq.~(\ref{ubtd2_13}) is:
\bqa
\koef&=&V\chi(\mw^2,s)N^2 N^{-}_{(3,2)}, \nll
%N^{-}_{(3,2)}&=&\sqrt{\frac{\smt\smb}{2}},\nll
\kk &=& 2\stb-\cpl\smt,\nll
\kkk&=& 2\stb\mtp+\cpl\mbt\smt.
\eqa
%Common to Eq.~(\ref{ubtd1_13}) and Eq.~(\ref{ubtd2_13}) are:
%\bqa
%\stb &=& s + \mtp\mbt,  \qquad
% \mtb\;=\; \mtp+\mbt, \nll
% \smt &=& s - \mtp^2, \qquad\qquad\;\, \smb \;=\; s - \mbt^2,
%\eqa
%and
%\bqa
%N    &=&\sqrt{\frac{2s}{\stb^2\cmi + s\mtp^2\cpl}}\,,
%N^{+}&=&\frac{\mtb}{\sqrt{2}},
%N^{-}&=&\frac{\stb}{\sqrt{2s}}.
%\eqa
% id Npuu(1,4)^2 = (mtp + mbt)^2/2;
% id Nmuu(1,4)^2 = (s + mbt*mtp)^2/2/s;
% id Nmuu(2,3)^2 = (s - mbt^2)*(s-mtp^2)/2/s;
% id Npvv(4,1)^2 = (mtp + mbt)^2/2;
% id Nmvv(4,1)^2 = (s + mbt*mtp)^2/2/s;
% id Nmvv(3,2)^2 = (s - mbt^2)*(s - mtp^2)/2/s;

Next we give HAs for the $t$ channel process of annihilation of different
flavors; they are simpler than the previous ones.

$\bullet$ {Helicity amplitudes for $b\bar{d}\to t\bar{u}$
%\_1\_14
%-------
\bqa
\label{ubtd1_14}
\cal{H}_{-++-} &=&  \koef\cpl\bigl({\cal F}_{LL}-\mtp {\cal F}_{LD}-\mbt {\cal F}_{RD}\bigr),
\nll
\cal{H}_{-+++} &=& -\koef\sqrt{s}\sin\vartheta\left(\frac{\mtp}{s} {\cal F}_{LL}\right.
\nll &&
  \left. -{\cal F}_{LD}-\frac{\mtp\mbt}{s} {\cal F}_{RD} \right),
\nll
\cal{H}_{+++-} &=&  \koef\sqrt{s}\sin\vartheta
  \left[\frac{\mbt}{s} \left( {\cal F}_{LL}  -\mtp {\cal F}_{LD}\right)-{\cal F}_{RD} \right],
\nll
\cal{H}_{++++} &=& -\koef\Bigl[ \frac{\mtp\mbt\cmi}{s} {\cal F}_{LL} -2{\cal F}_{RL}
\nll &&
             +\cpl\left(\mbt {\cal F}_{LD}+ \mtp {\cal F}_{RD}\right) \Bigr].
\eqa
%---

$\bullet$ {Helicity amplitudes for $\bar{b}d\to\bar{t}u$
%\_2\_14
%-------
\bqa
\label{ubtd2_14}
\cal{H}_{----} &=& -k_{0} \Bigl[\frac{\mtp\mbt\cmi}{s} {\cal F}_{LL} - 2 {\cal F}_{RL}
\nll &&
+\cpl \left(\mtp {\cal F}_{LD}+\mbt {\cal F}_{RD} \right)\Bigr],
\nll
\cal{H}_{---+} &=& -k_{0}\sqrt{s}\sin\vartheta\left(\frac{\mbt}{s} {\cal F}_{LL} \right.
\nll &&
                    \left.   -{\cal F}_{LD}-\frac{\mtp\mbt}{s} {\cal F}_{RD} \right),
\nll
\cal{H}_{+---} &=&  k_{0} \sqrt{s}\sin\vartheta
\nll &&
      \times\left[ \frac{\mtp}{s} ({\cal F}_{LL}-\mbt{\cal F}_{LD})-{\cal F}_{RD} \right],
\nll
\cal{H}_{+--+} &=&  k_{0}\cpl\bigl({\cal F}_{LL}-\mbt {\cal F}_{LD}-\mtp {\cal F}_{RD}\bigr).
\eqa
In Eqs.~(\ref{ubtd1_14}) and~(\ref{ubtd2_14})
\bq
\koef=V\chi(\mw^2,s) P^{+}(s,0,\mbt) P^{+}(s,0,\mtp)\,,
\eq
with a function typical to helicity amplitudes:
\bq
P^{\pm}(s,x,y)=\sqrt{s-(x\pm y)^2}\,.
\label{P_function}
\eq
%--
}

\section{Infrared regularization by the complex top quark mass.}
%-----------------------------------------------------------------------
Infrared divergences associated with interactions of photons with on-mass-shell top legs
in the limit of zero width of the top quark may be regularized by any conventional method
(photon mass, dimensional regularisation), but physically they are naturally
regularized by the presence of the finite top quark width $\Gamma_t$.
Here this approach will be exploited.

%------------------------------------------------------
\subsection{Top quark width and virtual QED correction}
%------------------------------------------------------

There are three diagrams contributing to virtual one-loop QED corrections
involving the top quark: see Figs.~\ref{Diagram_tdec_vac-virt1},
\ref{Diagram_tdec_vac-virt2} and \ref{Diagram_tdec_vac-virt3}.

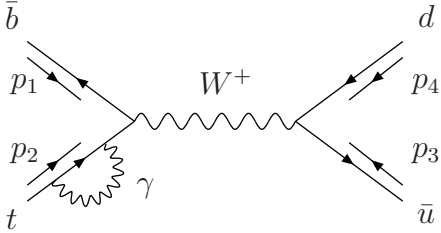
\begin{figure}[!ht]
\begin{center}
   \begin{picture}(125,80)(210,10)
%    \GOval(270,40)(34,5)(90){0.02}
     \Photon(240,40)(300,40){3}{6}
     \PhotonArc(222,22)(12,200,52){2}{10}
     \Text(275,55)[]{\large $W^+$}
     \Text(245,15)[]{\large $\gamma$}

     \ArrowLine(240,40)(200,70)
     \ArrowLine(200,10)(240,40)
     \ArrowLine(340,70)(300,40)
     \ArrowLine(300,40)(340,10)

     \ArrowLine(200,16)(220,32)
     \ArrowLine(200,64)(220,48)
     \ArrowLine(340,16)(320,32)
     \ArrowLine(340,64)(320,48)

     \Text(200,54)[]{\large $p_1$}
     \Text(200,28)[]{\large $p_2$}
     \Text(350,28)[]{\large $p_3$}
     \Text(350,54)[]{\large $p_4$}

     \Text(195,80)[]{\large$\bar{b}$}
     \Text(195,3)[]{\large $t$}
     \Text(350,5)[]{\large $\bar u$}
     \Text(350,80)[]{\large$d$}

   \end{picture}
\end{center}
\caption[The self-energy photonic QED correction for $t \bar{b} \bar{u} d\to 0$ processes]
        {The self-energy photonic QED correction for $t \bar{b} \bar{u} d\to 0$ processes}
\label{Diagram_tdec_vac-virt1}
\end{figure}

\begin{figure}[!ht]
\begin{center}
   \begin{picture}(125,80)(210,10)
%    \GOval(270,40)(34,5)(90){0.02}
     \Photon(240,40)(300,40){3}{6}
     \PhotonArc(240,40)(12,144,216){2}{5}
     \Text(275,55)[]{\large $W^+$}
     \Text(220,40)[]{\large $\gamma$}

     \ArrowLine(240,40)(200,70)
     \ArrowLine(200,10)(240,40)
     \ArrowLine(340,70)(300,40)
     \ArrowLine(300,40)(340,10)

     \ArrowLine(200,16)(220,32)
     \ArrowLine(200,64)(220,48)
     \ArrowLine(340,16)(320,32)
     \ArrowLine(340,64)(320,48)

     \Text(200,54)[]{\large $p_1$}
     \Text(200,28)[]{\large $p_2$}
     \Text(350,28)[]{\large $p_3$}
     \Text(350,54)[]{\large $p_4$}

     \Text(195,80)[]{\large$\bar{b}$}
     \Text(195,3)[]{\large $t$}
     \Text(350,5)[]{\large $\bar u$}
     \Text(350,80)[]{\large$d$}
   \end{picture}
\end{center}
\caption[The vertex photonic QED correction]
        {The vertex photonic QED correction}
\label{Diagram_tdec_vac-virt2}
\end{figure}
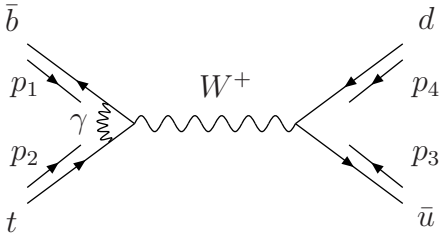

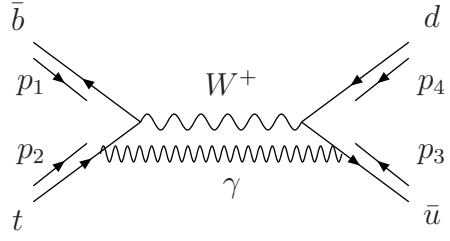
\begin{figure}[!ht]
\begin{center}
   \begin{picture}(125,80)(210,10)
%    \GOval(270,40)(34,5)(90){0.02}
     \Photon(240,40)(300,40){3}{6}
     \Photon(225,28)(315,28){3}{20}
     \Text(275,55)[]{\large $W^+$}
     \Text(275,15)[]{\large $\gamma$}

     \ArrowLine(240,40)(200,70)
     \ArrowLine(200,10)(240,40)
     \ArrowLine(340,70)(300,40)
     \ArrowLine(300,40)(340,10)

     \ArrowLine(200,16)(220,32)
     \ArrowLine(200,64)(220,48)
     \ArrowLine(340,16)(320,32)
     \ArrowLine(340,64)(320,48)

     \Text(200,54)[]{\large $p_1$}
     \Text(200,28)[]{\large $p_2$}
     \Text(350,28)[]{\large $p_3$}
     \Text(350,54)[]{\large $p_4$}

     \Text(195,80)[]{\large$\bar{b}$}
     \Text(195,3)[]{\large $t$}
     \Text(350,5)[]{\large $\bar u$}
     \Text(350,80)[]{\large$d$}
   \end{picture}
\end{center}
\caption[The direct box photonic QED correction]
        {The direct box photonic QED correction}
\label{Diagram_tdec_vac-virt3}
\end{figure}

The appropriate
formalism was developed in Ref.~\cite{Bardin:2009wv}. % and \cite{Bardin:2009ix}.
In particular, it was shown that of the four scalar
form factors ${\cal{F}}$ only ${\cal{F}}_{LL}$ contains IRD QED as well as weak
contributions which can be separated thus:
\bqa
{\cal{F}}_{LL}
=1+\frac{e^2}{16\pi^2}\tilde{\cal{F}}_{LL}^{\,\rm QED}
+\frac{g^2}{16\pi^2}\tilde{\cal{F}}_{LL}^{\,\rm weak}
\eqa
and the expression for
${\displaystyle \tilde{\cal{F}}_{LL}^{\,\rm QED}}$
was explicitly expressed in terms of standard Passarino--Veltman (PV)
functions $C_0$.
To regulate the amplitude the photon was given a mass $\lambda$. However, one
can alternatively introduce the top quark width, i.e. instead of
$$
C_0\left(-\mtp^2,\,-\mbt^2,\,Q^2;\,\mtp,\,\lambda,\,\mbt\right)
$$
we now write
\bqa
C_0\left(-\mtp^2,\,-\mbt^2,\,Q^2;\,\mtpt,\,0,\,\mbt\right)
\eqa
with
%metka1
%% ${\displaystyle \tilde{m}_t^2=\mtp^2-i\,\mtp\Gamma_t}$.
   ${\displaystyle
 \mtpt^2=\mtp^2+\Delta_t\,,\quad\Delta_t\;=\;-i\mtp\wtp\,}$.

 The corresponding replacement for the IRD term due to
 Fig.~\ref{Diagram_tdec_vac-virt1} reads
\bq
\ln\left(\frac{\mtp^2}{\lambda^2}\right)
 \to -2\frac{\mtpt^2}{\mtp^2}\ln\left(\frac{-i\mtp\wtp}{\mtpt^2}\right).
\eq
% 2(1-\mtp^2 B_{0p}(-\mtp^2;0,\mtpt)).

The expression for the amplitude of Fig.~\ref{Diagram_tdec_vac-virt3} contains
an infrared and mass singular $D_0$ function.
In Ref.~\cite{Bardin:2009ix} we have shown that these functions can be reduced
by standard Passarino--Veltman reduction to an infrared finite mass singular
auxiliary function ${\displaystyle J^d_{AW}}$ and a mass singular $C_0$
function.
This has become standard \SANC approach which we apply here.

The relevant $C_0$ function is
%---------------------------------------
%%
%% $\bullet$  $C_0\left(-\mtp^2,\,-\mbt^2,\,Q^2;\,\mtpt,\,0,\,\mbt\right)$
%%
\bqa
 C_0^{\rm IR} &=& C_0(-\mtp^2,-\mbt^2,\Qs;\mtpt,0,\mbt)
\\
         &=& \frac{1}{\Sd}\Biggl\{
          \ln\left(\ydl\right)\ln\left(1-\frac{1}{\ydl}\right)
          -\Litwo\left(\frac{1}{\ydl}\right)
\nll &&\quad\;
          -\Litwo\left(\frac{1-\ydl}{\yll-\ydl}\right)
          +\Litwo\left( -\frac{\ydl}{\yll-\ydl}\right)
\nll &&\quad\;
          -\Litwo\left(\frac{1-\ydl}{\ylll-\ydl}\right)
          +\Litwo\left( -\frac{\ydl}{\ylll-\ydl}\right)
\nll &&\quad\;
          -\ln\left(\ydll\right)\ln\left(1-\frac{1}{\ydll}\right)
          +\Litwo\left(\frac{1}{\ydll}\right)
\nll &&\quad\;
          +\Litwo\left(\frac{1-\ydll}{\yll-\ydll}\right)
          -\Litwo\left( -\frac{\ydll}{\yll-\ydll}\right)
\nll &&\quad\;
          +\Litwo\left(\frac{1-\ydll}{\ylll-\ydll}\right)
          -\Litwo\left( -\frac{\ydll}{\ylll-\ydll}\right)
                     \Biggr\},
\nonumber
\eqa
where
\bqa
      \Bd  &=&\Qs+\mtp^2-\mbt^2+\Qs/|\Qs|\ieps,
\nll
      \Sd  &=&\sqrt{(\Qs)^2+2\Qs(\mtp^2+\mbt^2)+(\mtp^2-\mbt^2)^2},
\nll
      \ydl &=&\frac{\Bd-\Sd}{2\Qs}\,,\qquad
      \ydll\;=\;\frac{\Bd+\Sd}{2\Qs}\,,
\eqa
and
\bqa
      \Bl  &=&\Qs+\mtpt^2-\mbt^2,
\nll
      \Sl  &=&\sqrt{(\Qs)^2+2\Qs(\mtpt^2+\mbt^2)+(\mtpt^2-\mbt^2)^2},
\nll
      \yll &=&\frac{\Bl-\Sl}{2\Qs}\,,\qquad
      \ylll\;=\;\frac{\Bl+\Sl}{2\Qs}\,.
\eqa

\noindent
%%
%%$\bullet$
%%
 Its limit for $\mbt\to 0$ is
\bqa
C_0^{\rm IR} &=& \frac{1}{\Bd}\Biggl\{
  \ln\left(\frac{\Delta_t}{\Bl}\right)\ln\left(\frac{\mbt^2}{\mtp^2}\right)
%\nll &&
           -2\ln\left(\frac{\Bd}{\mtp^2}\right)\ln\left(\frac{\Delta_t}{\mtp^2}\right)
\nll &&
+\left[ \frac{1}{2} \ln\left(\frac{\Bd}{\mtp^2}\right)
+\ln\left(\frac{\Bl}{\mtp^2}\right) \right]
               \ln\left(\frac{\Bd}{\mtp^2}\right)
\nll &&
           -\Litwo\left(-\frac{\Delta_t}{\mtp^2}\right)
           +\Litwo\left(-\frac{\Delta_t}{\Bd}  \right)
\nll &&
           -\Litwo\left( \frac{\Delta_t}{\Bl}\right)
           -\Litwo\left( \frac{Q^2}{\Bl}\right)
           +\zeta(2)      \Biggr\}.
\eqa
Here
\bqa
      \Bl &=& Q^2+\mtpt^2,
\nll
      \Bd &=& Q^2+\mtp^2-i\varepsilon.
\eqa
%---

%-----------------------------------------------------
\subsection{Top quark width and hard photon radiation}
%-----------------------------------------------------

We must also consider the amplitude with real photon emission off the $t$
quark, see Fig.~\ref{Diagram_tdec_vac-real}. This amplitude
\begin{figure}[!ht]
\begin{center}
   \begin{picture}(125,80)(210,10)
%    \GOval(270,40)(34,5)(90){0.02}
     \Photon(240,40)(300,40){3}{6}
     \Photon(225,28)(255,15){3}{7}

     \Text(275,55)[]{\large $W^+$}
     \Text(262,12)[]{\large $\gamma$}

     \ArrowLine(240,40)(200,70)
     \ArrowLine(200,10)(240,40)
     \ArrowLine(340,70)(300,40)
     \ArrowLine(300,40)(340,10)

     \ArrowLine(200,16)(220,32)
     \ArrowLine(200,64)(220,48)
     \ArrowLine(340,16)(320,32)
     \ArrowLine(340,64)(320,48)

     \Text(200,54)[]{\large $p_1$}
     \Text(200,28)[]{\large $p_2$}
     \Text(350,28)[]{\large $p_3$}
     \Text(350,54)[]{\large $p_4$}

     \Text(195,80)[]{\large$\bar{b}$}
     \Text(195,3)[]{\large $t$}
     \Text(350,5)[]{\large $\bar u$}
     \Text(350,80)[]{\large$d$}
   \end{picture}
\end{center}
\caption[The top Bremsstrahlung diagram]
        {The top Bremsstrahlung diagram}
\label{Diagram_tdec_vac-real}
\end{figure}
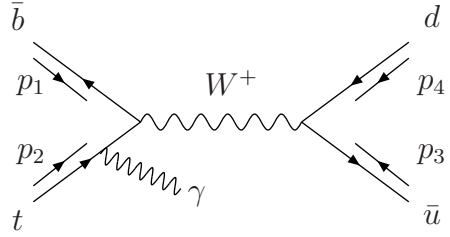
involves a propagator with denominator\\
${\displaystyle \left(p_2-p_5\right)^2+m_t^2-i\,m_t\,\Gamma_t}$
which evaluates to
${\displaystyle z_2-i\,m_t\,\Gamma_t}$ with
$z_2=-2p_2\cdot p_5$.

The amplitude of the hard photon radiation process without taking account of
the top quark width can be presented in the following form:
\bqa
\cA^{(0)}_{\gamma} = \cA^{(0)}_{t} + \cA_{udbW}.
\label{tamp0}
\eqa
Here $\cA^{(0)}_{t}$ is the amplitude of the radiating top quark and
$\cA_{udbW}$ is the rest. The amplitude (\ref{tamp0}) is gauge
invariant: $\cA^{(0)}_{\gamma} \cdot p_5 = 0$.

One can define the amplitude of the hard photon radiation process
taking account of the top quark width as
\bqa
\cA^{}_{\gamma} = \cA^{}_{t} + \cA_{udbW} =
\frac{z_2}{z_2-i \mtp \wtp}\cA^{(0)}_{t} + \cA_{udbW},\;\;
\label{tamp}
\eqa
where the top quark propagator with  width $\wtp$ was introduced.

The amplitude (\ref{tamp}) destroys gauge invariance:\linebreak
$\cA_{\gamma} \cdot p_5 \neq 0$.
The gauge violation is of order $\Gamma_t/m_t$.

To recover gauge invariance one can modify the amplitude
(\ref{tamp}) in the following way:
\bqa
\cA^{}_{\gamma}&\to&\cA^{\prime}_{\gamma}
=\frac{z_2}{z_2-i \mtp \wtp}\cA^{(0)}_{\gamma}
\nll
&=&
\frac{z_2}{z_2-i \mtp \wtp}\left[\cA^{(0)}_{t} + \cA_{udbW}\right].
\label{tamp-over}
\eqa

Such transformation (called {\em overall scheme, (o)}) obviously leads to gauge
invariance and the mod-squared amplitude will have the form
\bqa
\left|\cA^{over}_{\gamma}\right|^2&=&
\frac{z_2^2}{z_2^2 + \mtp^2 \wtp^2}\biggl\{ \left|\cA^{(0)}_{t}\right|^2
\nll
&+& \left[\left(\cA^{(0)}_{t}\right)^{*}\cA_{udbW}
+\cA^{(0)}_{t}\left(\cA_{udbW}\right)^{*}\right]
\nll
&+&\left|\cA_{udbW}\right|^2 \biggr\}.
\label{tamp-over2}
\eqa

As is seen from equation (\ref{tamp-over2}), all terms are now
regularized by the top quark width but the last term is multiplied by
 an {\em artificial} factor which cannot be treated in the
soft (Born-like) kinematics.

To avoid the problem with integration one can introduce a new
{\em gauge non-invariant} amplitude (which we call
 {\em fixed$_{1}$  scheme (f$_{1})$}):

The amplitude  $\cA^{}_{\gamma}$ with
 top quark width introduced {\it ab initio}~(\ref{tamp})
is called {\em fixed$_1$}:
\bqa
\left|\cA^{fixed_1}_{\gamma}\right|^2&=&
\frac{z_2^2}{z_2^2 + \mtp^2 \wtp^2} \left|\cA^{(0)}_{t}\right|^2
\nll
&+& 2\Re\frac{z_2}{z_2 - i\mtp \wtp}\cA^{(0)}_{t}\left(\cA_{udbW}\right)^{*}
\nll
&+& \left|\cA_{udbW}\right|^2.
\label{tamp-fix1}
\eqa
An alternative version, called {\em fixed$_2$},
obtains by setting equal to one the coefficient of
 $\left|\cA_{udbW}\right|^2$ in Eq.~(\ref{tamp-over2}):
\vspace*{-8mm}

\bqa
\left|\cA^{fixed_2}_{\gamma}\right|^2&=&
\frac{z_2^2}{z_2^2 + \mtp^2 \wtp^2}\biggl\{  \left|\cA^{(0)}_{t}\right|^2
\nll
&+&\left[\left(\cA^{(0)}_{t}\right)^{*}\cA_{udbW}
+\cA^{(0)}_{t}\left(\cA_{udbW}\right)^{*}\right]\biggr\}
\nll
&+& \left|\cA_{udbW}\right|^2.
\label{tamp-fix2}
\eqa
%---
\vspace*{-11mm}

In the next chapter we present numerical results for all width
schemes introduced above in order to estimate the numerical
impact of gauge violation in {\em fixed schemes}.
\vspace*{-5mm}

\section{Numerical Results\label{ResAndComp}}
%--------------------------------------------
In this section we present the \SANC results for the cross sections of single
top quark production and for the decay widths of the top quark. The tree level
contributions, both Born and single real photon emission, are compared with
{\tt CompHEP}~\cite{Pukhov:1999gg}.
 All numerical results for this section were produced
with the standard {\tt SANC INPUT} working in the $\alpha(0)$ EW scheme,
if not stated otherwise.

The standard {\tt SANC} input parameters set reads:
\vspace*{-10mm}

\bq
\begin{array}[b]{lcllcllcl}
G_F & = & 1.16637 \times 10^{-5} \GeV^{-2}, & & & \\
\alpha(0) &=& 1/137.035999, &
\alpha_s &=& 0.00729735257, \\
\mw & = & 80.403 \GeV, & \gw & = & 2.141  \GeV, \\
\mz & = & 91.1876\GeV, & \gz & = & 2.4952 \GeV, \\
\mh & = & 120\GeV,     & m_t & = & 174.2\;\GeV, \\
m_u & = & 62\;\MeV,    & m_d & = & 83\;\MeV,  \\
m_c & = & 1.5\;\GeV,   & m_s & = & 215\;\MeV, \\
m_t & = & 174.2\;\GeV, & m_b & = & 4.7\;\GeV, \\
|V_{ud}| &=&1, & |V_{cs}| &=& 1, \\
|V_{us}| &=&0, & |V_{cd}| &=& 0, \\ 
|V_{tb}| &=&1. & & &
\vspace*{-10mm}
\label{input}
\end{array}
\eq
%--
This section is subdivided into three subsections: for the decay channel,
 $s$ channel and $t$ channel;
 for the latter subprocesses results for both
{\it direct} ($bu\to td$) and {\it crossed} ($b\bar{d}\to t\bar{u}$)
 channels are presented separately.

Each subsection, in turn, consists of two paragraphs.
The first paragraphs contain a comparison with {\tt Comp-} {\tt HEP} of the
Born cross section and of the hard photon contribution with a cut on
photon energy $E_\gamma$ in the rest frame of the top (for decay channel)
 and in the cms of the initial particles for $s$ and $t$ channels.

In the second paragraphs we present a study of complete EWRC at the partonic level.
We show the decay widths for the decay channel and cross sections for the
reactions at the tree and one-loop levels. The one-loop EWRC is defined as usual:
\bq
\delta=\frac{{\cal{O}}^{\rm 1loop}-{\cal{O}}^{\rm tree}}{{\cal{O}}^{\rm tree}}\,,
\eq
%--
where ${\cal{O}}$ is either width ($\Gamma$) or cross section ($\sigma$).

Furthermore, the dimensionless soft-hard separator $\bar{\omega}$ subdivides
soft and hard real photon contributions. Their sum does not depend on
$\bar{\omega}$.

For decays $a\to b+c+d+\gamma$ with $s=-(p_a-p_b)^2$ and $s'=M^2_{c,d}$ and
for the processes, $a+b\to c+d+\gamma$ with $s=M^2_{a,b}$ and $s'=M^2_{c,d}$
\bq
\bar{\omega}=1-\frac{s'}{s}, \qquad \mbox{and}
\quad E_\gamma \geq \frac{s}{2}\bar{\omega}\,.
\eq
%--
For each of the four channels we present Tables illustrating the existence of
an $E_\gamma$ ($\bar{\omega}$) stability plateau for two options $\Gamma_{t}=0,\neq0$, 
and answering the question how
big the difference is of the EWRC between the two $\Gamma_{t}$ options.
For reactions, we also investigate the question of the initial quark masses ($m_q$) 
independence of the subtracted quantities.

For the explanation of the necessity of the subtraction procedure and its
realization in the $\MSbar$ scheme  within {\tt SANC} we refer to our earlier
papers~\cite{Arbuzov:2005dd}--\cite{Arbuzov:2007db}.

\subsection{Decay Channel $t\to b + u + \bar{d}$}
%------------------------------------------------

\subsubsection{SANC--CompHEP comparison}
%---------------------------------------

No new comparison for this channel was done compared to
Ref.~\cite{Bardin:2009wv}. If the top quark width is taken into account,
{\tt SANC} and {\tt CompHEP} numbers agree well.
% [[A comparison with CompHEP might be added in few days.]]

\subsubsection{One-loop EW corrections}
%--------------------------------------

Process $t\to b + \mu^+ + \bar{\nu}_\mu\,.$

\begin{table}[!h]
\begin{center}
\begin{tabular}{|l|l|l|l|}
\hline
\multicolumn{4}{|c|}{$\Gamma^{\rm Born}$=0.1490949(2) GeV}\\ \hline
\multicolumn{4}{|c|}{$\Gamma_{t}=0$}                      \\ \hline
$E_\gamma$, GeV  &  $10^{-1}$  & $10^{-2}$   & $10^{-3}$  \\ \hline
$\Gamma^{\rm 1loop}$& 0.159503(1) & 0.159495(2) & 0.159499(7)\\ \hline
$\delta        $& 6.981(1)    & 6.975(1)    & 6.978(5)   \\ \hline
\multicolumn{4}{|c|}{$\Gamma_{t}\neq0$, {\em fixed}$_1$}  \\ \hline
$\Gamma^{\rm 1loop}$& 0.160736(2) & 0.160784(2) & 0.160787(8)\\ \hline
$\delta        $& 7.810(1)    & 7.841(2)    & 7.842(5)   \\ \hline
\end{tabular}
\end{center}
\caption[]{
The total lowest-order and one-loop corrected widths $\Gamma^{\rm Born}$
and $\Gamma^{\rm 1l}$ in GeV and relative one-loop correction $\delta$
in $\%$ for $E_\gamma = 10^{-1},10^{-2},10^{-3}$~GeV.\label{Table11}}
\end{table}
%====================================================

As is seen from Table~\ref{Table11}, there is good $\bar{\omega}$ stability for
$E_\gamma \leq 10^{-2}$~GeV and about +0.9\% increase of the EWRC for this
partial decay width if one uses $\Gamma_{t}\neq0$ option
as compared to the $\Gamma_{t}=0$ one.

\noindent
Process $t\to b + u + \bar{d}\,$.

\begin{table}[!h]
\begin{center}
\begin{tabular}{|l|l|l|l|}
\hline
\multicolumn{4}{|c|}{$\Gamma^{\rm Born}$=0.4472847(7) GeV}  \\ \hline
\multicolumn{4}{|c|}{$\Gamma_{t}=0$}                       \\ \hline
$E_\gamma$, GeV     &  $10^{-1}$  & $10^{-2}$    & $10^{-3}$ \\ \hline
$\Gamma^{\rm 1loop}$ & 0.47922(1)  & 0.47920(1)  & 0.47918(1)\\ \hline
$\delta$            & 7.139(1)    & 7.135(1)    & 7.131(3)  \\ \hline
\multicolumn{4}{|c|}{$\Gamma_{t}\neq0$, {\em fixed}$_1$}     \\ \hline
$\Gamma^{\rm 1loop}$ & 0.48293(1)  & 0.48312(1) & 0.48311(1) \\ \hline
$\delta$            & 7.969(1)    & 8.012(1)   & 8.010(2)   \\ \hline
\end{tabular}
\end{center}
\caption{The same but for $t\to b + u + \bar{d}$ process.\label{Table12}}
\end{table}
%----------

As is seen from Table~\ref{Table12}, one can draw similar conclusions for
both considered partial decay channels.

\subsection{s channel}
%---------------------
There are two generic $s$ channel processes:
\bq
\bar{d}+u\to\bar{b}+t,\quad\mbox{and}\quad\bar{u}+d\to\bar{t}+b.
\eq

\subsubsection{SANC--CompHEP comparison}
%---------------------------------------
It is sufficient to consider only one: $\bar{u}+d\to\bar{t}+b.$
For this comparison we use the {\tt CompHEP (v.4.5.1)} setup with
a cut on the cms photon energy $E_{\gamma}\geq$ 2 GeV.
\begin{table}[!h]
\begin{center}
\begin{tabular}{|l|c|l|l|l|}
\hline
$\sqrt{\hs}/GeV$&ws& 200 & 1000      & 7000    \\ \hline
\multicolumn{5}{|c|}{$\Gamma_{t}=0$}           \\ \hline
CompHEP&     & 8.172(1) & 17.19(1) & 0.7966(2) \\ \hline
SANC   &     & 8.173(1) & 17.19(1) & 0.7964(1) \\ \hline
\multicolumn{5}{|c|}{$\Gamma_{t}=1.54688$ GeV} \\ \hline
CompHEP&  o  & 7.835(1) & 16.93(1) & 0.7882(1) \\ \hline
       &  f  & 7.788(1) & 16.90(1) & 0.7884(1) \\ \hline
SANC   &  o  & 7.834(1) & 16.93(1) & 0.7885(1) \\ \hline
       &f$_1$& 7.788(1) & 16.91(1) & 0.7884(1) \\ \hline
       &f$_2$& 7.787(1) & 16.91(1) & 0.7884(1) \\ \hline
\end{tabular}
\end{center}
\caption[]{Comparison of the cross section
$\sigma^{\rm{hard}}(\sqrt{\hs},\bar{\omega})$ (fb)
for three cms energies, two options: $\Gamma_{t}=0,\neq 0$,
and three {\it width schemes (ws): o-overall, f-fixed}
(f$_{1(2)}$={\it fixed}$_{1(2)}$ for the case of {\tt SANC}).\label{Table20}}
\end{table}
%----------

As is seen from Table~\ref{Table20}, there is good agreement of
numbers obtained from \SANC and {\tt CompHEP} within the statistical
errors for all considered options and cms energies.
It is worth to note that the two versions of width schemes agree well for
all energies except
near the threshold and that the {\em fixed}$_1$ scheme agrees with
the CompHEP {\em fixed} scheme as expected.

\subsubsection{One-loop EW corrections}
%--------------------------------------
The numbers of this subsection are produced with
{\tt SANC} setup, Eq.~(\ref{input}).
The aim is to demonstrate the intervals of stability
of one-loop corrected EW cross sections
$\sigma^{\rm 1loop}$ and relative EWRC $\delta$ against variation of the soft-hard separation
parameter, $\bar{\omega}$ and to study the difference between the two options:
$\Gamma_{t}=0,\neq0$.
\begin{table}[!h]
\begin{center}
\begin{tabular}{|l|l|l|l|}
\hline
\multicolumn{4}{|c|}{$\sqrt{\hs}$=200 GeV, $\sigma^{\rm Born}$=0.30809055(1) pb}\\ \hline
\multicolumn{4}{|c|}{$\Gamma_{t}=0$}                             \\ \hline
 $\bar{\omega}$           &$10^{-4}$   & $10^{-5}$  & $10^{-6}$   \\ \hline
 $\hat{\sigma}^{\rm 1loop}$    & 0.315169(2)& 0.315167(3)& 0.315167(4)\\ \hline
 $\delta$                 & 2.297(1)   & 2.297(1)   & 2.297(1)   \\ \hline
\multicolumn{4}{|c|}{$\Gamma_{t}\neq0$, {\em fixed}$_1$}           \\ \hline
 $\hat{\sigma}^{\rm 1loop}$    & 0.315469(2)& 0.315471(3)& 0.315472(3)\\ \hline
 $\delta$                 & 2.395(1)   & 2.396(1)   & 2.396(1)   \\ \hline
%\multicolumn{4}{|c|}{$\Gamma_{t}\neq0$, {\em fixed}_2}            \\ \hline
% $\hat{\sigma}^{1l}$      & 0.315469(2)& 0.315471(3)& 0.315471(3) \\ \hline
% $\delta^{1l}$            & 2.395(1)   & 2.395(1)   & 2.395(1)    \\ \hline
\end{tabular}
\end{center}
\caption[]{The total lowest-order cross sections $\hat{\sigma}^{\rm Born}$,
the one-loop corrected cross sections $\hat{\sigma}^{\rm 1loop}$ in pb and
relative corrections $\delta$ in $\%$ at three values of
$\bar{\omega}=10^{-4},10^{-5},10^{-6}$ and two of $\Gamma_{t}$.\label{Table21}}
\end{table}
%----------

\begin{table}[!h]
\begin{center}
\begin{tabular}{|l|l|l|l|}
\hline
\multicolumn{4}{|c|}{$\sqrt{\hs}$=1000 GeV, $\sigma^{\rm Born}$=0.105977185(2) pb}\\ \hline
\multicolumn{4}{|c|}{$\Gamma_{t}=0$}                              \\ \hline
 $\bar{\omega}$           &  $10^{-4}$  & $10^{-5}$   & $10^{-6}$   \\ \hline
 $\hat{\sigma}^{\rm 1loop}$& 0.102872(3) & 0.102874(3) & 0.102875(4)\\ \hline
 $\delta$                 &-2.930(2)    &-2.928(3)    &-2.927(4)   \\ \hline
\multicolumn{4}{|c|}{$\Gamma_{t}\neq0$, {\em fixed}$_1$}           \\ \hline
 $\hat{\sigma}^{\rm 1loop}$& 0.102753(3) & 0.102775(4) & 0.102774(5)\\ \hline
 $\delta$                 &-3.043(3)    &-3.021(3)    &-3.023(4)   \\ \hline
%\multicolumn{4}{|c|}{$\Gamma_{t}\neq0$, {\em fixed_2}              \\ \hline
% $\hat{\sigma}_1$        & 0.102753(3) & 0.102775(4) & 0.102774(5) \\ \hline
% $\delta$                &-3.043(3)    &-3.021(4)    &-3.023(4)    \\ \hline
\end{tabular}
\end{center}
\caption{The same but for $\sqrt{\hs}$=1000 GeV.\label{Table22}}
\end{table}
%----------

\begin{table}[!h]
\begin{center}
\begin{tabular}{|l|l|l|l|l|}
\hline
\multicolumn{4}{|c|}{$\sqrt{\hs}$=7000 GeV, $\sigma^{\rm Born}$=2.23529503(4) fb}\\ \hline
\multicolumn{4}{|c|}{$\Gamma_{t}=0$}                                            \\ \hline
 $\bar{\omega}$           & $10^{-5}$ & $10^{-6}$ & $10^{-7}$   \\ \hline
 $\hat{\sigma}^{\rm 1loop}$& 1.5917(1) & 1.5917(2) & 1.5919(1)  \\ \hline
 $\delta$                 & -28.79(1) & -28.80(1) & -28.78(1)  \\ \hline
\multicolumn{4}{|c|}{$\Gamma_{t}\neq0$, {\em fixed}$_1$}        \\ \hline
 $\hat{\sigma}^{\rm 1loop}$& 1.5815(1) & 1.5895(2) & 1.5895(2)  \\ \hline
 $\delta$                 & -29.25(1) & -28.89(1) & -28.89(1)  \\ \hline
\end{tabular}
\end{center}
\caption{The same but for $\sqrt{\hs}$=7000 GeV and for $\bar{\omega}=10^{-5},10^{-6},10^{-7}$
and the cross sections given in fb.
\label{Table23}}
\end{table}
%----------

Tables~\ref{Table21}--\ref{Table23} were computed with three values of
$\bar{\omega}$ and for two options $\Gamma_{t}=0,\neq0$.
As far as $\bar{\omega}$-stability is concerned, we see that it depends on
the $\Gamma_{t}$-option.
For $\Gamma_{t}=0$ there is stability for all considered $\sqrt{\hs}$
and $\bar{\omega}$.
For $\Gamma_{t}\neq0$ it depends on the cms energy $\sqrt{\hs}$:
for 200 GeV stability sets in already at $\bar{\omega}=10^{-4}$, for 1000 GeV
at $\bar{\omega}=10^{-5}$
and for 7000 GeV at $\bar{\omega}=10^{-6}$.

\begin{table}[!h]
\begin{center}
\begin{tabular}{|l|l|l|l|}
\hline
\multicolumn{4}{|c|}{$\sqrt{\hs}$=200 GeV, $\Gamma_{t}=0$}       \\ \hline
 $\bar{\omega}$           &$10^{-4}$   & $10^{-5}$   & $10^{-6}$  \\ \hline
 $\hat{\sigma}^{\MSbar}_1$ & 0.328077(2)& 0.328082(4)& 0.328083(4)\\ \hline
 $\delta^{\MSbar}_1$       & 6.487(1)   & 6.489(1)   & 6.489(1)   \\ \hline
 $\hat{\sigma}^{\MSbar}_2$ & 0.328073(2)& 0.328078(2)& 0.328076(3)\\ \hline
 $\delta^{\MSbar}_2$       & 6.486(1)   & 6.488(1)   & 6.487(1)   \\ \hline
\end{tabular}
\end{center}
\caption{
The total one-loop corrected $\MSbar$ subtracted quantities
$\hat{\sigma}^{\MSbar}$ in pb and corresponding $\delta^{\MSbar}$ in $\%$ at
$\bar{\omega} = 10^{-4},10^{-5},10^{-6}$, respectively. Subscript $1$ means
that light quark masses are used as in Eq.~(\ref{input}),
 while $2$ means that 10 times smaller masses are used.   \label{Table24}}
\end{table}
%----------

However, one can see that the difference between the two options, although it
persists, is of order 1 per mille in absolute deviation,
which is well below any reasonable estimate of the theoretical uncertainty.
We therefore conclude that one may use the usual infrared regularization for 
$s$ channel processes.

The numbers of Tables~\ref{Table24}--\ref{Table26} were produced with the aim
to demonstrate the independence from light quark masses of the subtracted
quantities, $\hat{\sigma}^{\MSbar}$ and $\delta^{\MSbar}$ for $\Gamma_{t}=0$
option at three cms energies. One sees that the light quark mass independence
holds in all considered cases at the level much lower than 1 per mille,
which is quite sufficient for practical applications.

\begin{table}[!h]
\begin{center}
\begin{tabular}{|l|l|l|l|}
\hline
\multicolumn{4}{|c|}{$\sqrt{\hs}$=1000 GeV, $\Gamma_{t}=0$}        \\ \hline
 $\bar{\omega}$           &$10^{-4}$    & $10^{-5}$    & $10^{-6}$   \\ \hline
 $\hat{\sigma}^{\MSbar}_1$ & 0.100084(3) & 0.100086(3) & 0.100087(4)\\ \hline
 $\delta^{\MSbar}_1$       & -5.561(2)   & -5.559(3)   &-5.558(4)   \\ \hline
 $\hat{\sigma}^{\MSbar}_2$ & 0.100084(3) & 0.100086(3) & 0.100087(4)\\ \hline
 $\delta^{\MSbar}_2$       & -5.565(4)   & -5.567(5)   & -5.568(7)  \\ \hline
\end{tabular}
\end{center}
\caption{The same but for $\sqrt{\hs}$=1000 GeV.
\label{Table25}}
\end{table}
%----------

\begin{table}[!h]
\begin{center}
\begin{tabular}{|l|l|l|l|}
\hline
\multicolumn{4}{|c|}{$\sqrt{\hs}$=7000 GeV, $\Gamma_{t}=0$}    \\ \hline
 $\bar{\omega}$           & $10^{-5}$ & $10^{-6}$  & $10^{-7}$  \\ \hline
 $\hat{\sigma}^{\MSbar}_1$ & 1.4566(1) & 1.4565(2) & 1.4568(1)  \\ \hline
 $\delta^{\MSbar}_1$       & -34.84(1) & -34.84(1) & -34.83(1)  \\ \hline
 $\hat{\sigma}^{\MSbar}_2$ & 1.4569(2) & 1.4570(2) & 1.4570(2)  \\ \hline
 $\delta^{\MSbar}_2$       & -34.82(1) & -34.82(1) & -34.82(1)  \\ \hline
\end{tabular}
\end{center}
\caption{The same but for $\sqrt{\hs}$=7000 GeV
but for cross sections given in fb 
and for $\bar{\omega} = 10^{-5},10^{-6},10^{-7}$.
\label{Table26}}
\end{table}
%----------

\subsection{$t$ channel}
%-----------------------

\subsubsection{SANC--CompHEP comparison}
%---------------------------------------

This is the most complicated case: $t$ channel cross sections usually show up bad
statistical convergence. For this comparison we use the {\tt CompHEP (v.4.5.1)} setup,
but with non-zero masses of the $u$ and $d$ quarks
(accessed via $bc\to ts\gamma$ channel).
For the Tables of this subsection we used $m_u=m_d=m_q= 66$ MeV,
and $10m_q$ means $m_u=m_d= 660$ MeV. The cut on the cms photon energy was $E_{\gamma}\geq$ 2 GeV.

\begin{table}[!h]
\begin{center}
\begin{tabular}{|l|c|c|c|c|}
\hline
$\sqrt{\hs}/GeV$&ws& 200   & 1000     & 7000      \\ \hline
\multicolumn{5}{|c|}{$\Gamma_{t}=0$, $m_q$}       \\ \hline
CompHEP&       & 0.2468(1) & 9.078(1) & 18.68(1)  \\ \hline
SANC(S)&       & 0.2467(1) & 9.090(1) & 18.45(1)  \\ \hline
\multicolumn{5}{|c|}{$\Gamma_{t}\neq 0$, $m_q$}   \\ \hline
CompHEP& f     & 0.2392(1) & 9.161(1) & 18.68(1)  \\ \hline
SANC(S)& f$_1$ & 0.2392(1) & 9.174(1) & 18.70(1)  \\ \hline
\multicolumn{5}{|c|}{$\Gamma_{t}=0$, $10m_q$}     \\ \hline
CompHEP&       & 0.1740(1) & 7.512(1) & 16.11(1)  \\ \hline
SANC(S)&       & 0.1745(1) & 7.514(1) & 16.12(1)  \\ \hline
SANC(F)&       & 0.1740(1) &  ---     &  ---      \\ \hline
\multicolumn{5}{|c|}{$\Gamma_{t}\neq 0$, $10m_q$} \\ \hline
CompHEP& f     & 0.1694(1) & 7.597(1) & 16.36(1)  \\ \hline
SANC(S)& f$_1$ & 0.1698(1) & 7.599(1) & 16.37(1)  \\ \hline
SANC(F)& f$_1$ & 0.1694(1) &  ---     &  ---      \\ \hline
\end{tabular}
\end{center}
\caption[]{Comparison of the cross section $\sigma^{\rm{hard}}(\sqrt{\hs},\bar{\omega})$, fb
for the process $b+u\to t+d$ for three cms energies;
four options: $(\Gamma_{t}=0,\neq 0)\otimes(m_q,10m_q)$;
and {\it fixed width scheme (ws):}
(\it{f$_{1}$=fixed}$_1$ for the case of {\tt SANC}).
\label{Table31}}
\end{table}
%----------

Furthermore, rows marked ``SANC(S)'' were computed retaining $m_q$ or $10m_q$
only in fermion propagators radiating a photon, while ``SANC(F)''
means that light quark masses were kept everywhere (``F''ully massive case).
Table~\ref{Table32} contains the same information as
Table~\ref{Table31} but for the process $b+\bar{d}\to t+\bar{u}.$

\begin{table}[!h]
\begin{center}
\begin{tabular}{|l|c|c|c|c|}
\hline
$\sqrt{\hs}/GeV$&ws& 200   & 1000     & 7000     \\ \hline
\multicolumn{5}{|c|}{$\Gamma_{t}=0$, $m_q$}       \\ \hline
CompHEP&       & 0.1562(1) & 8.620(1) & 18.27(1) \\ \hline
SANC   &       & 0.1562(2) & 8.634(1) & 18.33(1) \\ \hline
\multicolumn{5}{|c|}{$\Gamma_{t}\neq 0$, $m_q$}  \\ \hline
CompHEP& f     & 0.1522(1) & 8.706(1) & 18.53(1) \\ \hline
SANC   & f$_1$ & 0.1522(1) & 8.717(1) & 18.57(1) \\ \hline
\multicolumn{5}{|c|}{$\Gamma_{t}=0$, $10m_q$}    \\ \hline
CompHEP&       & 0.1032(1) & 7.120(1) & 15.98(1) \\ \hline
SANC   &       & 0.1037(1) & 7.123(1) & 15.98(1) \\ \hline
SANC(F)&       & 0.1032(1) &  ---     & ---      \\ \hline
\multicolumn{5}{|c|}{$\Gamma_{t}\neq 0$, $10m_q$}\\ \hline
CompHEP& f     & 0.1009(1) & 7.204(1) & 16.22(1) \\ \hline
SANC   & f$_1$ & 0.1014(1) & 7.207(1) & 16.24(1) \\ \hline
SANC(F)& f$_1$ & 0.1009(1) &  ---     & ---      \\ \hline
\end{tabular}
\end{center}
\caption[]{The same comparison of the cross section $\sigma^{\rm hard}(\sqrt{\hs},\bar{\omega}), fb$
but for the process $b+\bar{d}\to t+\bar{u}.$
\label{Table32}}
\end{table}
%----------
Let us discuss the results of Tables~\ref{Table31}--\ref{Table32} which are
qualitatively the same, so we need not refer to the Table number.

%metka:2.8.10

Two options $(\Gamma_{t}=0,\neq 0)\otimes(m_q)$ (4 upper rows) show very
good agreement at the
threshold, $\sqrt{\hs}=200\,\mbox{GeV}$, and a notable deviation at TeV energies.
We assumed that this is
due to the mass singular origin of external light quark lines emitting photons,
which obviously is
getting worse at high energies. In this connection, we emphasize that
a special procedure of
numerical stabilization of terms giving rise to light quark mass
singularities was applied in
{\tt SANC}.

To check this assumption we considered options
$(\Gamma_{t}=0,\neq 0)\otimes(10m_q)$
and we immediately see the inverse trend: only satisfactory agreement at
$\sqrt{\hs}=200\,\mbox{GeV}$ and much better agreement at
TeV energies. This exercise convinced us that in the mass singular stable
regime (large masses) {\tt CompHEP} and \SANC results do agree.
It is easy to understand why with the latter options the situation near the
threshold is not so ideal. In \SANC we want to neglect terms with power
dependence on light quark masses, keeping only those that would lead to mass
singularities. Relatively large masses,
$10m_q$, start to matter near the threshold. (Note also that on passing to the
hadronic level
we will need to subtract mass quark singularities anyway: see next subsection).

%metka3.8.10 10 a.m.

To check this assumption we considered the fully massive variant {\tt SANC(F)} with options
$(\Gamma_{t}=0,\neq 0)\otimes(10m_q)$ and only at
$\sqrt{\hs}=200\,\mbox{GeV}$. As is seen from
the {\tt SANC(F)} rows of the Tables, the good agreement with {\tt CompHEP} is recovered.

Although {\tt SANC(F)} is identical to {\tt CompHEP}, we prefer to use the much faster {\tt SANC(S)}
version of the code since the masses of $u$, $d$ and $s$ quarks are very small.
The only exception is the $c$ quark whose mass effect will be obviously
notable near the threshold. Since we posses the {\tt SANC(F)} version
of the code, we can use a mixed variant with only $c$ quark mass dependence retained.

To conclude this section, we note that the results for the charge conjugate
channels are identical to those we have considered:
\bqa
\bar{b}+\bar{u}\to \bar{t}+\bar{d} \quad \mbox{to} \quad b+u\to t+d\,,
\nll
\bar{b}+d\to \bar{t}+u \quad \mbox{to} \quad b+\bar{d}\to t+\bar{u}\,.
\eqa
%---

%db-27
\subsubsection{One-loop EW corrections, process $bu\to td$\label{tdirect}}
%-------------------------------------------------------------------------
The numerical results for this subsection were again produced with the
{\tt SANC} setup, Eq.~(\ref{input}), with
the same aim to demonstrate the stabilility of one-loop corrected EW cross sections
$\sigma^{\rm 1loop}$ and relative EWRC $\delta$ against variation of the
soft-hard separator
$\bar{\omega}$ and to study the difference between the two options: $\Gamma_{t}=0,\neq0$.
\begin{table}[!h]
\begin{center}
\begin{tabular}{|l|l|l|l|}
\hline
\multicolumn{4}{|c|}{$\sqrt{\hs}$=200 GeV, $\sigma^{\rm Born}$=7.3551155(1) pb}\\ \hline
\multicolumn{4}{|c|}{$\Gamma_{t}=0$}                             \\ \hline
 $\bar{\omega}$           &$10^{-4}$   & $10^{-5}$   & $10^{-6}$  \\ \hline
 $\hat{\sigma}^{\rm 1loop}$       & 7.81964(2) & 7.81954(4) & 7.81956(4)\\ \hline
 $\delta$             & 6.3156(3)  & 6.3144(5)  & 6.3146(6) \\ \hline
\multicolumn{4}{|c|}{$\Gamma_{t}\neq0$, {\em fixed}$_1$}           \\ \hline
 $\hat{\sigma}^{\rm 1loop}$       & 7.81954(2) & 7.81956(3) & 7.81957(4)\\ \hline
 $\delta$             & 6.3143(3)  & 6.3145(4)  & 6.3147(5) \\ \hline
\end{tabular}
\end{center}
\caption[]{
The total lowest-order cross sections $\hat{\sigma}^{\rm Born}$,
the one-loop corrected cross sections $\hat{\sigma}^{\rm 1loop}$ in pb and
relative one-loop correction $\delta$ in $\%$ at three values of
$\bar{\omega}=10^{-4},10^{-5},10^{-6}$ and two values of $\Gamma_{t}$.\label{Table33}}
\end{table}
%----------

\begin{table}[!h]
\begin{center}
\begin{tabular}{|l|l|l|l|}
\hline
\multicolumn{4}{|c|}{$\sqrt{\hs}$=1000 GeV, $\sigma^{\rm Born}$=48.99340951(6) pb}\\ \hline
\multicolumn{4}{|c|}{$\Gamma_{t}=0$}                              \\ \hline
 $\bar{\omega}$            &$10^{-5}$   & $10^{-6}$  & $10^{-7}$   \\ \hline
 $\hat{\sigma}^{\rm 1loop}$       & 53.344(1). & 53.345(1). & 53.346(1). \\ \hline
 $\delta$             & 8.880(1).  & 8.881(2).  & 8.883(2).  \\ \hline
\multicolumn{4}{|c|}{$\Gamma_{t}\neq0$, {\em fixed}$_1$}            \\ \hline
 $\hat{\sigma}^{\rm 1loop}$       & 53.292(1)  & 53.293(1)  & 53.293(1)  \\ \hline
 $\delta$             & 8.773(1)   & 8.776(2)   & 8.776(2)   \\ \hline
\end{tabular}
\end{center}
\caption[]{The same but for $\sqrt{\hs}$=1000 GeV
and for $\bar{\omega} = 10^{-5},10^{-6},10^{-7}$.\label{Table34}}
\end{table}
%----------

\begin{table}[!h]
\begin{center}
\begin{tabular}{|l|l|l|l|}
\hline
\multicolumn{4}{|c|}{$\sqrt{\hs}$=7000 GeV, $\sigma^{\rm Born}$=50.82423111(8) pb}\\ \hline
\multicolumn{4}{|c|}{$\Gamma_{t}=0$}                             \\ \hline
 $\bar{\omega}$           &$10^{-6}$   & $10^{-7}$  & $10^{-8}$   \\ \hline
 $\hat{\sigma}^{\rm 1loop}$       & 55.696(2) & 55.697(2)  & 55.697(2)  \\ \hline
 $\delta$             &  9.586(3) &  9.588(4)  &  9.587(4)  \\ \hline
\multicolumn{4}{|c|}{$\Gamma_{t}\neq0$, {\em fixed}$_1$}         \\ \hline
 $\hat{\sigma}^{\rm 1loop}$       & 55.613(1) & 55.639(2)  & 55.640(2)  \\ \hline
 $\delta$             &  9.428(3) &  9.474(3)  &  9.476(4)  \\ \hline
\end{tabular}
\end{center}
\caption[]{The same but for $\sqrt{\hs}$=7000 GeV
and for $\bar{\omega} = 10^{-6},10^{-7},10^{-8}$.\label{Table35}}
\end{table}
%----------

Tables~\ref{Table33}--\ref{Table35} were computed with three values of
$\bar{\omega}$ and for two options $\Gamma_{t}=0,\neq0$.
As is seen, the intervals of $\bar{\omega}$-stability depend on $\sqrt{\hs}$.
The higher the cms energy,
the lower are the values of $\bar{\omega}$ required to reach stability.
For $\sqrt{\hs}$=200\,GeV, stability starts at $\bar{\omega}=10^{-4}$,
for $\sqrt{\hs}$=1000\,GeV
at $\bar{\omega}=10^{-5}$ and for $\sqrt{\hs}$=7000\,TeV it depends additionally
on the $\Gamma_{t}$-option.
For $\Gamma_{t}=0$ stability starts at $\bar{\omega}=10^{-6}$ but for
$\Gamma_{t}\neq0$ only at $\bar{\omega}=10^{-7}$.
However, one can see that the difference between the two options, although
it persists, is well below 1 per mille in absolute deviation
at lower energies, reaching 1 per mille only
at $\sqrt{\hs}$=7000\,GeV, that is the difference under study is below any
reasonable estimate of the theoretical uncertainty.
We therefore conclude that one may use the usual infrared
regularization also for $t$ channel processes.\\

\leftline{\em Subtraction of quark mass singularities}

\begin{table}[!h]
\begin{center}
\begin{tabular}{|l|l|l|l|}
\hline
\multicolumn{4}{|c|}{$\sqrt{\hs}$=200 GeV, $\sigma^{\rm Born}$=7.3551155(1) pb}\\ \hline
\multicolumn{4}{|c|}{$\Gamma_{t}=0$}                             \\ \hline
 $\bar{\omega}$           &$10^{-4}$   & $10^{-5}$   & $10^{-6}$  \\ \hline
 $\hat{\sigma}^{\MSbar}_1$ & 8.09370(2) & 8.09377(4) & 8.09380(4) \\ \hline
 $\delta^{\MSbar}_1$       & 10.042(1)  & 10.043(1)  & 10.043(1)  \\ \hline
 $\hat{\sigma}^{\MSbar}_2$ & 8.09116(4) & 8.09127(6) & 8.09129(8) \\ \hline
 $\delta^{\MSbar}_2$       & 10.043(1)  & 10.045(1)  & 10.045(1)  \\ \hline
\end{tabular}
\end{center}
\caption[]{
The total one-loop corrected $\MSbar$ subtracted quantities (see text)
$\hat{\sigma}^{\MSbar}$ in pb and corresponding $\delta^{\MSbar}$ in $\%$ at
$\bar{\omega} = 10^{-4},10^{-5},10^{-6}$.
Sub-index $1$ means that light quark masses used are as in Eq.~(\ref{input}),
while $2$ means that 10 times lower masses are used.
\label{Table36}}
\end{table}
%----------

The numbers of the Tables~\ref{Table36}--\ref{Table38} were produced with the aim
to demonstrate the light quark mass independence of the one-loop subtracted quantities
$\hat{\sigma}^{\MSbar}$ and $\delta^{\MSbar}$ for the $\Gamma_{t}=0$ option at
three cms energies and three values of $\bar{\omega}$.

\begin{table}[!h]
\begin{center}
\begin{tabular}{|l|l|l|l|}
\hline
\multicolumn{4}{|c|}{$\sqrt{\hs}$=1000 GeV, $\sigma^{\rm Born}$=48.99340953(6) pb}\\ \hline
\multicolumn{4}{|c|}{$\Gamma_{t}=0$}                             \\ \hline
 $\bar{\omega}$           &$10^{-5}$   & $10^{-6}$   & $10^{-7}$  \\ \hline
 $\hat{\sigma}^{\MSbar}_1$ & 53.416(1)  & 53.416(1)  & 53.417(1)  \\ \hline
 $\delta^{\MSbar}_1$       & 9.027(1)   & 9.028(2)   & 9.030(2)   \\ \hline
 $\hat{\sigma}^{\MSbar}_2$ & 53.416(1)  & 53.416(1)  & 53.417(2)  \\ \hline
 $\delta^{\MSbar}_2$       & 9.029(2)   & 9.029(2)   & 9.032(3)   \\ \hline
\end{tabular}
\end{center}
\caption[]{
The same but for $\sqrt{\hs}$=1000 GeV
and for $\bar{\omega} = 10^{-5},10^{-6},10^{-7}$.\label{Table37}}
\end{table}
%----------

\begin{table}[!h]
\begin{center}
\begin{tabular}{|l|l|l|l|}
\hline
\multicolumn{4}{|c|}{$\sqrt{\hs}$=7000 GeV, $\sigma^{\rm Born}$=50.82423111(8) pb}\\ \hline
\multicolumn{4}{|c|}{$\Gamma_{t}=0$}                             \\ \hline
 $\bar{\omega}$           &$10^{-6}$    & $10^{-7}$  & $10^{-8}$  \\ \hline
 $\hat{\sigma}^{\MSbar}_1$ & 55.699(2)  & 55.700(2)  & 55.699(2)  \\ \hline
 $\delta^{\MSbar}_1$       &  9.591(3)  &  9.593(4)  &  9.592(4)  \\ \hline
 $\hat{\sigma}^{\MSbar}_2$ & 55.698(2)  & 55.697(3)  & 55.696(3)  \\ \hline
 $\delta^{\MSbar}_2$       &  9.589(3)  &  9.588(5)  &  9.585(6)  \\ \hline
\end{tabular}
\end{center}
\caption[]{
The same but for $\sqrt{\hs}$=7000\,GeV 
and for $\bar{\omega} = 10^{-6},10^{76},10^{-8}$.\label{Table38}}
\end{table}
%----------

The stability of $\hat{\sigma}^{\MSbar}$ and $\delta^{\MSbar}$
seen in Tables~\ref{Table36}--\ref{Table38}
is a very important check of the correctness of
implementation of the complete
EWRC and of the initial quark mass singularity subtraction. The latter is
crucial at the envisaged stage of going to the hadron level,
i.e. convolution with parton density functions.

%db-28
\subsubsection{One-loop EW corrections: process $b\bar{d}\to t\bar{u}$}
%----------------------------------------------------------------------

Here we repeat the study of subsection~\ref{tdirect}, but now for the crossed
$t$ channel. Tables~\ref{Table39}--\ref{Table41}
 are analoguous to Tables~\ref{Table33}--\ref{Table35}.
They demonstrate the stabilility of $\sigma^{\rm 1l}$ and $\sigma^{\rm 1l}$
against variation of $\bar{\omega}$ and show the difference between
the options $\Gamma_{t}=0,\neq0$.

\begin{table}[!h]
\begin{center}
\begin{tabular}{|l|l|l|l|}
\hline
\multicolumn{4}{|c|}{$\sqrt{\hs}$=200 GeV, $\sigma^{\rm Born}$=4.495790646(3) pb}\\ \hline
\multicolumn{4}{|c|}{$\Gamma_{t}=0$}                               \\ \hline
 $\bar{\omega}$           &$10^{-4}$    & $10^{-5}$   & $10^{-6}$   \\ \hline
 $\hat{\sigma}^{\rm 1loop}$       & 4.87169(2) & 4.87165(2)  & 4.87165(3) \\ \hline
 $\delta$             & 8.3612(4)  & 8.3603(5)   & 8.3603(7)  \\ \hline
\multicolumn{4}{|c|}{$\Gamma_{t}\neq0$, {\em fixed}$_1$}             \\ \hline
 $\hat{\sigma}^{\rm 1loop}$       & 4.86989(5) & 4.86990(2)  & 4.86991(3) \\ \hline
 $\delta$             & 8.3212(4) 	& 8.3213(5)   & 8.3215(6)  \\ \hline
\end{tabular}
\end{center}
\caption[]{
The total lowest-order cross sections $\hat{\sigma}^{\rm Born}$,
the one-loop corrected cross sections $\hat{\sigma}^{\rm 1loop}$ in pb and
relative one-loop correction $\delta$ in $\%$ at three values of
$\bar{\omega}=10^{-4},10^{-5},10^{-6}$ and two values of $\Gamma_{t}$.\label{Table39}}
\end{table}
%----------

\begin{table}[!h]
\begin{center}
\begin{tabular}{|l|l|l|l|}
\hline
\multicolumn{4}{|c|}{$\sqrt{\hs}$=1000 GeV, $\sigma^{\rm Born}$=46.68695597(6) pb}\\ \hline
\multicolumn{4}{|c|}{$\Gamma_{t}=0$}                             \\ \hline
 $\bar{\omega}$           &$10^{-5}$    & $10^{-6}$  & $10^{-7}$  \\ \hline
 $\hat{\sigma}^{\rm 1loop}$       & 51.054(1)  & 51.054(1) & 51.055(1)  \\ \hline
 $\delta$             &  9.354(1)  &  9.354(2) &  9.356(2)  \\ \hline
\multicolumn{4}{|c|}{$\Gamma_{t}\neq0$, {\em fixed}$_1$}           \\ \hline
 $\hat{\sigma}^{\rm 1loop}$       & 51.004(1)  & 51.005(1) & 51.006(1)  \\ \hline
 $\delta$             &  9.246(1)  &  9.249(2) &  9.250(2)  \\ \hline
\end{tabular}
\end{center}
\caption[]{The same but for $\sqrt{\hs}$=1000 GeV
and for $\bar{\omega} = 10^{-5},10^{-6},10^{-7}$.\label{Table40}}
\end{table}
%----------

%\clearpage

\begin{table}[!h]
\begin{center}
\begin{tabular}{|l|l|l|l|}
\hline
\multicolumn{4}{|c|}{$\sqrt{\hs}$=7000 GeV, $\sigma^{\rm Born}$=50.72449055(7) pb}\\ \hline
\multicolumn{4}{|c|}{$\Gamma_{t}=0$}                           \\ \hline
 $\bar{\omega}$           &$10^{-6}$   & $10^{-7}$  & $10^{-8}$ \\ \hline
 $\hat{\sigma}^{\rm 1loop}$       & 55.581(2) & 55.583(2) & 55.580(2) \\ \hline
 $\delta$             & 9.575(3)  & 9.578(4)  & 9.572(4)  \\ \hline
\multicolumn{4}{|c|}{$\Gamma_{t}\neq0$, {\em fixed}$_1$}         \\ \hline
 $\hat{\sigma}^{\rm 1loop}$       & 55.495(2) & 55.527(2) & 55.528(2) \\ \hline
 $\delta$             & 9.404(3)  & 9.467(3)  & 9.469(4)  \\ \hline
\end{tabular}
\end{center}
\caption[]{The same but for $\sqrt{\hs}$=7000 GeV
and for $\bar{\omega} = 10^{-6},10^{-7},10^{-8}$.\label{Table41}}
\end{table}
%----------

 Tables~\ref{Table39}--\ref{Table41} show up properties very similar
to the Tables for the direct $t$ channel. Each Table shows the presence
of a plateau of stability in $\bar{\omega}$ and the same level of difference
between calculations for two options $\Gamma_{t}=0,\neq0$,
qualitatively with very similar energy dependence.

\leftline{\em Subtraction of quark mass singularities}

\begin{table}[!h]
\begin{center}
\begin{tabular}{|l|l|l|l|}
\hline
\multicolumn{4}{|c|}{$\sqrt{\hs}$=200 GeV, $\sigma^{\rm Born}$=4.495790646(3) pb}\\ \hline
\multicolumn{4}{|c|}{$\Gamma_{t}=0$}                             \\ \hline
 $\bar{\omega}$           &$10^{-4}$   & $10^{-5}$   & $10^{-6}$  \\ \hline
 $\hat{\sigma}^{\MSbar}_1$ & 4.92749(2) & 4.92749(2) & 4.92749(3) \\ \hline
 $\delta^{\MSbar}_1$       & 9.6023(4)  & 9.6023(5)  & 9.6023(7)  \\ \hline
 $\hat{\sigma}^{\MSbar}_2$ & 4.92515(3) & 4.92513(4) & 4.92514(5) \\ \hline
 $\delta^{\MSbar}_2$       & 9.6065(6)  & 9.6061(9)  & 9.606(1)   \\ \hline
\end{tabular}
\end{center}
\caption[]{
The total one-loop corrected $\MSbar$ subtracted quantities
$\hat{\sigma}^{\MSbar}$ in pb and corresponding $\delta^{\MSbar}$ in $\%$ at
$\bar{\omega} = 10^{-4},10^{-5},10^{-6}$. 
Sub-index $1$ means that light quark masses used are as in Eq.~(\ref{input}), 
while $2$ means that 10 times lower masses are used.
\label{Table42}}
\end{table}
%----------

The numbers of Tables~\ref{Table42}--\ref{Table44} were also produced with
the aim to demonstrate  the light quark mass independence of the subtracted
quantities $\hat{\sigma}^{\MSbar}$  and $\delta^{\MSbar}$
for option $\Gamma_{t}=0$ at three cms energies.

\begin{table}[!h]
\begin{center}
\begin{tabular}{|l|l|l|l|}
\hline
\multicolumn{4}{|c|}{$\sqrt{\hs}$=1000 GeV, $\sigma^{\rm Born}$=46.68695597(6) pb}\\ \hline
\multicolumn{4}{|c|}{$\Gamma_{t}=0$}                             \\ \hline
 $\bar{\omega}$           &$10^{-5}$   & $10^{-6}$  & $10^{-7}$  \\ \hline
 $\hat{\sigma}^{\MSbar}_1$ & 51.086(6) & 51.0858(8) & 51.087(1)  \\ \hline
 $\delta^{\MSbar}_1$       &  9.422(1) &  9.422(2)  &  9.424(2)  \\ \hline
 $\hat{\sigma}^{\MSbar}_2$ & 51.0856(9)& 51.086(1)  & 51.087(2)  \\ \hline
 $\delta^{\MSbar}_2$       &  9.424(2) &  9.424(3)  &  9.428(3)  \\ \hline
\end{tabular}
\end{center}
\caption[]{
The same but for $\sqrt{\hs}$=1000 GeV
and for $\bar{\omega} = 10^{-5},10^{-6},10^{-7}$.\label{Table43}}
\end{table}
%----------

\begin{table}[!h]
\begin{center}
\begin{tabular}{|l|l|l|l|}
\hline
\multicolumn{4}{|c|}{$\sqrt{\hs}$=7000 GeV, $\sigma^{\rm Born}$=50.72449055(7) pb}\\ \hline
\multicolumn{4}{|c|}{$\Gamma_{t}=0$}                           \\ \hline
 $\bar{\omega}$           &$10^{-6}$   & $10^{-7}$  & $10^{-8}$ \\ \hline
 $\hat{\sigma}^{\MSbar}_1$ & 55.583(2) & 55.585(2) & 55.582(2)  \\ \hline
 $\delta^{\MSbar}_1$       & 9.579(3)  & 9.582(4)  & 9.575(4)   \\ \hline
 $\hat{\sigma}^{\MSbar}_2$ & 55.583(2) & 55.585(3) & 55.576(3)  \\ \hline
 $\delta^{\MSbar}_2$       & 9.579(4)  & 9.582(5)  & 9.564(7)   \\ \hline
\end{tabular}
\end{center}
\caption[]{
The same but for $\sqrt{\hs}$=7000\,GeV
and for $\bar{\omega} = 10^{-6},10^{-7},10^{-8}$.\label{Table44}}
\end{table}
%----------

Again, Tables~\ref{Table42}--\ref{Table44}, although within larger statistical
errors, show the stability of $\hat{\sigma}^{\MSbar}$
and $\delta^{\MSbar}$ for all cms energies and $\bar{\omega}$ values
considered, confirming the correctness of implementation of the complete
EWRC also for the case of the crossed $t$ channel.

\section{Conclusions and Outlook\label{concl}}
%--------------------------------------------- 
In this paper we describe the implementation into the \SANC framework of
the complete one-loop EW
calculations, including hard bremsstrahlung contributions, for the processes
of the top quark decays a of $s$ and $t$ channel production, the latter two
at the partonic level,. The essentially new aspect of
this paper  is the study of regularisation of the top quark--photon infrared
divergences with aid of the complex mass of the top quark. For this reason,
we briefly come back to the top decay channels, since this issue was not
studied in our previous papers, Ref.~\cite{Arbuzov:2007ke,Bardin:2009wv}.
A comparison of these electroweak corrections computed within this new approach 
with those computed by the conventional method showed a sizable ($~1\%$)
effect for top decays and only per mille level effect for both considered
top production processes.
In this paper we limit ourselves to 4-fermion type of processes. For this reason we do not
consider $t$ quark production channel in association with $W$ boson. This channel will be 
presented elsewhere.

We have presented analytical expressions for the covariant amplitude of the
process and the helicity amplitudes for three different cross channels within
the standard \SANC multichannel approach, Ref~\cite{Bardin:2007wb}.
In this approach the one-loop covariant amplitude
is computed only once for to-the-vacuum-annihilation
processes, see Figs.~\ref{Diagram_tdec_vac} and~\ref{Diagram_atdec_vac}.
To get the CA for a physical channel a cross transformation
of external 4-momenta is performed.
The helicity amplitudes have been calculated for each
channel separately.

Within the \SANC framework, we have created the standard FORM and FORTRAN modules,
see,~Ref.\cite{Andonov:2008ga}, compiled into a package {\tt sanc\_cc\_v1.40} which
may be downloaded from \SANC project homepages.
All the calculations were done using a combination of analytic and Monte
Carlo integration methods which will make it easy to impose experimental
cuts in furtcoming calculations for the $pp$-collisions at LHC, which will
be the subject of a subsequent paper.

The emphasis of this paper is to be assured of the correctness of our results. We
observe the independence of the form factors on gauge parameters 
(all calculations were done in $R_{\xi}$ gauge), checked the stability of the result
against variation of the soft-hard separation parameter $\bar{\omega}$ and the
independence of the $\MSbar$ subtracted quantities off the initial quark
masses which is crucial for calculations at the hadronic level.

As has become \SANC standard, we tried to compare our numerical results for
these channels with other independent calculations. For the decay channels
it was done  in our preious papers (both EW~\cite{Arbuzov:2007ke} and
QCD~\cite{Andonov:2007zz}), showing good agreement.
As usual, the Born level and the hard photon contributions of all three
channels were checked against {\tt CompHEP} package and we found very
good agreement for both $\Gamma_{t}=0,\neq0$ options.

As far as the comparison of EWRC for the production processes is concerned,
the situation is not satisfactory. We did not find in the literature any
calculations for the $s$ channel. 
For the $t$ channel there are many papers, see, for instance, 
Refs.~\cite{Beccaria:2006ir,Beccaria:2007tc,Beccaria:2008av,Mirabella:2008gj}
and references therein. The most appropriate is Ref.~\cite{Beccaria:2006ir},
where however the hard photon contribution is not included. We tried to suppress it
in our calculations, but got only rough agreement with their Fig.~3(a).
The most advanced paper Ref.~\cite{Beccaria:2008av} contains
results at the hadronic level and, at present, we can not yet make a comparison.

The results presented in this papers lay a solid base for subsequent
extensions of calculations for the single top production channels at hadron colliders,
basically LHC.

\bigskip\noindent
{\bf Acknowledgements.}
This work is partly supported by Russian Foundation for Basic Research
grant $N^{o}$ 10-02-01030. 

The authors are grateful to Gizo Nanava for a useful discussion on the
subject of helicity amplitudes.

WvS is indebted to the directorate of the Dzhelepov Laboratory of Nuclear
Problems, JINR, Dubna, for the hospitality extended to him during June 2010.

\bibliographystyle{utphys_spires}
\addcontentsline{toc}{section}{\refname}\bibliography{Top_EW}
\end{document}